\documentclass[10pt]{article}%
\usepackage{graphicx}
\usepackage{times}
\usepackage{latexsym}
\usepackage{amsmath}%
\usepackage{amsfonts}%
\usepackage{amssymb}
\title{\bf GRAVITATION AS FIELD AND CURVATURE} 
\author{ L. V.
Verozub\\ Kharkov National University, 
\\E-mail: verozub@gravit.kharkov.ua}
\date{}
\begin{document}
\maketitle
\begin{abstract}
We argue that space-time properties are not absolute with respect to the
used frame of reference as is to be expected according to  ideas of 
relativity of
space
and time properties by Berkley -
Leibnitz - Mach - Poincar\'{e}.
From this point of view gravitation may manifests itself both as a field
in Minkowski space-time and as space-time curvature. 
If the motion of test particles is described by the Thirring Lagrangian,
then in the inertial frames of reference, where space-time is 
pseudo-Euclidean, gravitation manifests itself as a field. In reference 
frames,
whose reference body is formed by point masses moving under the effect of
the field,  it appears as Riemannian curvature
 which in these frames is other than zero.  
For realization of the idea the
author bimetric gravitation equations are considered. The spherically
- symmetric solution of the equations in Minkowski space-time does not
lead to the physical singularity in
the center. 
The energy of the gravitational field of a point mass is finite.
It follows from the properties of the gravitational force that
there can exist stable compact supermassive configurations of Fermi-gas 
without an events horizon.
\end{abstract}

\twocolumn

\section{Introduction}

The key reason preventing a correct inclusion of the Einstein  theory
of gravitation in the interactions unification is that gravity is
identified with space-time curvature. It is also a cause of such
unsolved problems of the theory as an operational definition of
the observable variables, the energy - momentum tensor problem and gravity
quantization. In the present paper starting from
\cite{Vr_Verozub81}, \cite{Vr_Verozub95} we consider a likely
reason of gravity geometrization. We argue that the gravitation
properties are not absolute with respect to the used frame of
reference. In inertial frames of reference gravitation can be
considered as a field in flat space-time, while in so called
proper frames of reference it manifests itself as space-time
curvature.

The author's gravitation equations which realize this idea are considered
in details. They do not contradict available experimental data. The  
physical consequences resulting from the equations differ very little from 
the
ones in general relativity if the distances from the attracting mass are
much
larger than the Schwarzschild radius $r_{g}$. However, they are completely 
different
at the distances equal to $r_{g}$ or less than that. 
A number of new physical consequences follow from the equations.


\section{Primary  Principles}

The geometrical properties of space-time can be described only by
means of measuring instruments. At the same time, the description
of the properties of measuring instruments, strictly speaking,
requires knowledge of space-time geometry. One of the implications
of it is that geometrical properties of space and time have no
experimentally verifiable significance by themselves but only
within the aggregate ''geometry + measuring instruments''. We got
aware of it owing to Poincar\'{e} \cite{Vr_Poincare}. It is a
development of the idea 
by Berkley - Leibnitz - Mach 
about relativity of space-time properties
which is an alternative to the well known Newtonian approach.

If we proceed from the conception of relativity of space-time, we 
assume that there is no way of quantitative description of
physical phenomena other than attributing them to a certain frame
of reference which in itself is a physical device for space and
time measurements. But then the relativity of the geometrical
properties of space and time mentioned above is nothing else but
relativity of space-time geometry with respect to the frame of
reference being used.\footnote[1]{There is an important difference
between a frame of reference (as a physical device) and a
coordinate system (as way to parameterize points of space-time)
\cite{Vr_Rodichev}}.

Thus, it should be assumed that the concept of the reference frame as a
physical object, whose properties are given and are independent of the 
properties
of space and time, is approximate, and only the aggregate ''frame of 
reference
+ space-time geometry'' has a sense.

The Einstein theory of gravitation demonstrates relativity of
space-time with respect to distribution of matter. However,
space-time relativity with respect to measurement instruments
hitherto has not been realized in physical theory. An attempt to
show that there is also space-time relativity to the used reference frames
for the first time  has
been undertaken in \cite{Vr_Verozub81}, \cite{Vr_Verozub95}.

At present we do not know how the space-time geometry in inertial
frames of reference (IFRs) is connected with the frames
properties. Under the circumstances, we simply postulate
(according to special relativity) that space - time in IFRs is
pseudo-Euclidean. Next, we find a space-time metric differential
form in noninertial frames of reference (NIFRs) from the viewpoint
of an observer in the NIFR who proceeds from the relativity of
space and time in the Berkley - Leibnitz - Mach - Poincar\'{e} (BLMP) sense. 
It is shown that there are 
reasons to believe that side by side with generally accepted
viewpoint on motion in noninertial frames of reference as a
relative motion there can also be another point of view. According to
this viewpoint the metric differential form $ds$ in the NIFR is
completely conditioned by the properties of the frame being used
as is to be expected according to the idea of relativity of space
and time in the BLMP sense.

\section{The Metric Form $ds$ in NIFR.}


By a noninertial frame of reference we mean the frame, whose body
of reference is fo/-rm/-ed by the point masses moving in the IFR under
the effect of a given force field.

It would be a mistake to identify "a priori" a transition from
the IFR to the NIFR with the transformation of coordinates related
to the frames. If we act in such a way, we already assume that the
properties of the space-time in both frames are identical.
However, for an observer in the NIFR, who proceeds from the
relativity of space and time in the BLMP sense, space-time
geometry is not given "a priori" and must be ascertained from the
analysis of experimental data.

We shall suppose that the reference body (RB) of the IFR or NIFR
is formed by the identical point masses $m_{p}$. If the observer is at
rest in one of the frames, his world line will coincide with the
world line of some point of the reference body. It is obvious to
the observer in the IFR that the accelerations of the point masses
forming the reference body are equal to zero. Of course, this fact
occurs in relativistic sense too. That is, if the differential
metric form of space-time in the IFR is denoted by $d\eta$ and
$\nu^{\alpha} = dx^{\alpha}/d\eta$ is the 4- velocity vector of
the point masses forming the reference body, then the absolute
derivative of the vector $\nu^{\alpha}$ is equal to zero,i.e.
\begin{equation}
\label{Vr_3/1 }D\nu^{\alpha} /d \eta= 0 .
\end{equation}
(We mean that an arbitrary coordinate system is used).

Does this fact occur for an observer in the NIFR ? That is, if
the differential metric form of space-time in the NIFR is denoted
by $ds$, does the 4-velocity vector $\zeta^{\alpha} = dx^{\alpha}
/ds$  of the point masses forming the reference body of this NIFR
obey the equation
\begin{equation}
\label{Vr_3/2}D \zeta^{\alpha} /ds = 0
\end{equation}

? The answer depends on whether space and time are absolute in
Newtonian sense or they are relative in the BLMP sense.

If space and time are absolute, the point masses of the NIFR
reference body for an observer in this NIFR are at relative rest.
A notion of relative acceleration can be determined in a covariant
way \cite{Vr_Dehnen}. This value is equal to zero. However,
eqs.(\ref{Vr_3/2}), strictly speaking, are not satisfied.

If space and time are relative in the BLMP sense, then for
observers in the IFR and NIFR the motion of the point masses
forming the reference body, which are kinematically
equivalent, must be dynamically equivalent too (both in the
nonrelativistic and relativistic sense). That is, if from the
viewpoint of the observer in the IFR, the point masses forming the
NIFR RB are at rest ( are not subject to the influence of forces
),then from the viewpoint of the observer in the NIFR the
point masses forming the RB of his frame are at rest too (are not
subject to the influence of forces either ). In other words, if
for the observer in the IFR the world lines of the IFR RB points
are, according to eq. (\ref{Vr_3/1 }), the geodesic lines, then for
the observer  in the NIFR the world lines of the NIFR RB points
are also the geodesic lines in his space-time, which can be
expressed by eq. (\ref{Vr_3/2}). The differential equations of
these world lines at the same time are the Lagrange equations of
motion of the NIFR RB points. The Lagrange equations, describing
the motion of the identical RB point masses in the IFR, can be
obtained from the Lagrange action $S$ by the principle of least action.
Therefore, the equations of the geodesic lines can be obtained
from the differential metric form $ds=k\ dS$, where $k$ is the
constant, $dS=L(x,\overset{.}{ x})dt$ and $L$ is the Lagrange function. 
The constant $k=-(m_{p} c)^{-1}$, as it follows from the
analysis of the case when the frame of reference is inertial. It
is equal to $-(m_{p} c)^{-1}$.

Thus, if we proceed from relativity of space and time in the BLMP sense, 
then
the differential metric form of space-time in the NIFR can be expected to 
have
the following form
\begin{equation}
\label{Vr_5/4}ds = -(m_{p} c)^{-1} \; dS(x,dx).
\end{equation}
In this equation $S$ is the Lagrangian action describing (in an IFR) the
motion of the identical point masses $m_{p}$, forming the NIFR
reference body.

So, the properties of space-time in the NIFR are entirely determined by the
properties of the used frame in accordance with the idea of relativity of
space and time in the BLMP sense.

Consider two examples of the NIFR.

1.The reference body is formed by noninteracting electric charges 
moving in a constant homogeneous electric field $E$. 
The motion of the charges is described in 
Cartesian coordinates by the Lagrangian
\begin{equation}
L = - m_{p} c^{2}\ (1-V^{2} /c^{2} )^{1/2} +E~e~ x ,
\label{Vr_lagrframe1}%
\end{equation}
where $V$ is the speed of the charges.

According to eq. 
(\ref{Vr_lagrframe1}) the space - time metric differential form in
this frame is
given by%

\begin{equation}
\label{Vr_5/6}ds = d\eta- (w x/c^{2} ) dx^{0},
\end{equation}
where $d\eta= (c^{2} dt^{2} - dx^{2} - dy^{2} - dz^{2} )^{1/2} $,
is the metric differential form of the pseudo -
Euclidean space - time in the IFR and $w=eE/m$ is the acceleration 
of the charges.

2. The reference body consists of noninteracting electric charges
in a constant homogeneous magnetic field $H$ directed along the
axis $z$. The Lagrangian describing the motion of the
particles can be written as follows \cite{Vr_Landau}
\begin{equation}
\label{Vr_5/7}L = - m_{p} c^{2} (1-V^{2} /c^{2} )^{1/2} - (m_{p} \Omega_{0}
/2)(\overset{\cdot}{x}y - x\overset{\cdot}{y}) ,
\end{equation}
where $\overset{\cdot}{x} =dx/dt$, $\overset{\cdot}{y} = dx/dt$ 
and $\Omega_{0}=eH/2mc$.

The points of such a system rotate in the plane $xy$ around the axis $z$ 
with
the angular frequency
\begin{equation}
\omega=\Omega_{0} [1+ (\Omega_{0} r /c )^{2}]^{-1/2}     
\label{Vr_omega (r)}
\end{equation}
where $r=(x^{2} +y^{2})^{1/2}$.
The linear velocities of the BR points tend to $c$ when $r \to\infty$.  

 For the given NIFR
\begin{equation}
\label{6/8}ds = d\eta+ (\Omega_{0} /(2c)\ (ydx - xdy).
\end{equation}

In the above NIFR $ds$ is of the form
\begin{equation}
ds = {\mathcal F}(x,dx) , \label{Vr_dsRanders}%
\end{equation}
where ${\mathcal F}(x,dx) = d\eta+ f_{\alpha}(x) dx^{\alpha} ,  f_{\alpha}$ 
is a
vector field and
\[
d\eta= [\eta_{\alpha\beta}(x)dx^{\alpha} dx^{\beta}]^{1/2}
\]
is the differential metric form of pseudo-Euclidean space-time of
the IFR in the used coordinate system. $\mathcal{F}$ is a homogeneous 
function of the first degree in
$dx^{\alpha}$.  Therefore, generally
speaking, the space-time in NIFR is Finslerian \cite{Vr_Rund} with
the sign-indefinite differential metric form (\ref{Vr_dsRanders}). 

\section{Space and Time Measurements in NIFR}


For the $3+1$ decomposition of space - time in noninertial frames
of reference to 3-space and time we proceed from a covariant
method which goes back to Ulman, Komar, Dehnen and other authors
\cite{Vr_Dehnen}. An ideal clock is a local periodic process
measuring the length of its own world line $\gamma$ to a certain
scale. For an observer in the NIFR the direction of time in the
point $x^{\alpha}$ is given by the vector of the 4-velocity
$\zeta^{\alpha}$ of the BR point.

The physical 3-space in each point is orthogonal to the vector $\zeta^{\alpha
}$. The arbitrary vector $\xi^{\alpha}$ in the point $x^{\alpha}$ can be
represented as follows:%

\begin{equation}
\xi^{\alpha} = \overline{\xi}^{\alpha} + \beta\zeta^{\alpha
},\label{Vr_3+1}%
\end{equation}
where $\overline{\xi}^{\alpha}$ are the spatial components and $\beta$ is 
the
function of $x^{\alpha}$.

Suppose any space vector $\overline{\xi}^{\alpha}$ in the
point $x^{\alpha}$ is orthogonal to the vector $\zeta^{\alpha}$ in
the sense of the Finslerian metric \cite{Vr_Rund}:
\begin{equation}
\overset{\ast}{\zeta}_{\alpha} \overline{\xi}^{\alpha} =0
,\label{Vr_ortogonality}%
\end{equation}
where $\overset{\ast}{\zeta}_{\alpha}$ is the covariant components
of the vector $\zeta^{\alpha}$, which are given by
\begin{equation}
\overset{\ast}{\zeta}_{\alpha}= \mathcal{F}(x, \zeta) 
\frac{\partial{\mathcal{F}(x,\zeta)}}
{\partial{\zeta^{\alpha}}}\label{Vr_covarvect}%
\end{equation}

Since $\mathcal{F}(x,\zeta^{\alpha}) = 1$ this vector is of the form
\begin{equation}
\overset{\ast}{\zeta_{\alpha}}= \eta_{\alpha\beta} \nu^{\alpha} + 
f_{\alpha} =
\nu_{\alpha}+ f_{\alpha},\label{Vr_decompositioncovectora}%
\end{equation}
where $\nu^{\alpha}= dx^{\alpha}/d\eta$ is the 4-velocity of the
reference body point in the IFR. By multiplying eq. (\ref{Vr_3+1})
to the vector $\overset{\ast}{\zeta_{\alpha}}$ we find that
$\beta= \overset{\ast}{\zeta }_{\alpha}\xi^{\alpha}$ and
\begin{equation}
\overline{\xi}^{\alpha}=H^{\alpha}_{\beta} \xi^{\beta},\label{Vr_spacecomp}%
\end{equation}
where $H^{\alpha}_{\beta}= \zeta^{\alpha}\overset{\ast}{\zeta}_{\beta}-
\delta^{\alpha}_{\beta}$ and $\delta^{\alpha}_{\beta}$ is the Kroneker delta.

Eq. (\ref{Vr_3+1}) yields for the vector
$\xi^{\alpha}=dx^{\alpha}$:
\begin{equation}
dx^{\alpha}=\overline{dx}^{\alpha} + c\ d\tau\zeta^{\alpha} ,
\label{Vr_3+1fordx}%
\end{equation}
where $\overline{dx}^{\alpha}$ is the spatial components of the vector
$dx^{\alpha}$ and
\begin{equation}
d\tau=c^{-1} \overset{\ast}{\zeta}_{\alpha}dx^{\alpha}\label{Vr_dT}%
\end{equation}
is the time element between the events in the points $x^{\alpha}$ and
$x^{\alpha}+dx^{\alpha}$ in the NIFR.

The metric form (\ref{Vr_dsRanders}) and the spatial projection of
the vector $dx^{\alpha}$ lead to the following covariant form of
the spatial element in the NIFR
\begin{equation}
dL= (- \eta_{\alpha\beta} \overline{dx}^{\alpha}\overline{dx}^{\beta} )^{1/2}
+ f_{\alpha}\ \overline{dx}^{\alpha}\label{Vr_dL}%
\end{equation}

This covariant equation is the simplest and clearest in the coordinates
system in which
\begin{equation}
\nu^{\alpha}=\delta_{0}^{\alpha}/(\eta_{00})^{1/2}\label{Vr_uinchroncoord}%
\end{equation}
Indeed, in eq. (\ref{Vr_dL})
\begin{equation}
-\eta_{\alpha\beta}\overline{dx}^{\alpha}\overline{dx}^{\beta}=H_{\alpha\beta
}dx^{\alpha}dx^{\beta},
\end{equation}
where
\begin{eqnarray}
H_{\alpha\beta}
=-\eta_{\alpha\beta}H_{\mu}^{\alpha}H_{\nu}^{\beta }\label{Vr_H}
=-\eta_{\alpha\beta}- \nonumber\\
\eta_{\mu\nu}\zeta^{\mu}\zeta^{\nu}\overset{\ast}%
{\zeta}_{\alpha}\overset{\ast}{\zeta_{\beta}}+\zeta_{\alpha}\overset{\ast
}{\zeta}_{\beta}+\zeta_{\beta}{\overset{\ast}{\zeta}_{\alpha}},
\end{eqnarray}
and $\zeta_{\alpha}=\eta_{\alpha\beta}\zeta^{\beta}$

In the used coordinates system
\begin{equation}
\zeta^{\alpha} = \nu^{\alpha} d\eta/ds = \lambda\delta^{\alpha}_{0}%
,\label{Vr_zetachroncoord}%
\end{equation}
where $\lambda=1/(\nu_{0} + f_{0})$ and $\nu_{0} = (\eta_{00})^{1/2}$. The
zero-component of the tensor $H_{\alpha\beta}$ is
\begin{equation}
H_{00}= g_{00} - 2 \eta_{00} \lambda(\nu_{0} + f_{0})+ (d\eta/ds)(\nu
_{0}+f_{0})^{2} .
\end{equation}
Since
\begin{equation}
d\eta/ds = (1 + f_{\alpha} \nu^{\alpha})^{-1} = (\eta_{00})^{1/2} \lambda
\end{equation}
and $\lambda(\nu_{0} + f_{0}) = 1$, the value of $H_{00}$ is equal to zero 
identically.

The components
\begin{multline}
H_{ik}=\eta_{ik} + 2 \lambda\eta_{i0} (\nu_{k}+f_{k}) +
\\(d\eta/ds)^{2}(\nu _{i}+f_{i})(u_{k}+ f_{k}),
\end{multline}
where $\nu_{i}=\eta_{0i}/ (\eta_{00})^{1/2}$. The spatial tensor $H_{ik}%
=\eta_{ik}$ with accuracy up to $V/c$, where $V$ is the linear speed of the
reference body points.

We have also
\begin{equation}
f_{\alpha} \overline{dx}^{\alpha} = \overline{f}_{\alpha} dx^{\alpha},
\end{equation}
where
\[
\overline{f}_{\alpha}= f_{\alpha} H^{\alpha}_{\beta}= f_{\alpha} - f_{0}
\lambda(\nu_{\alpha} +f_{\alpha}).
\]
The zero-component of the vector $\overline{f}_{\alpha}$ is equal to zero
identically and the spatial-components are equal to $f_{i}$ with accuracy up
to $V/c$.

Thus, in the used coordinates system the spatial element in the NIFR with
accuracy up to $V/c$ is of the form
\begin{equation}
dL=(-\eta_{ik}dx^{i}dx^{k})^{1/2} + f_{i}dx^{i}=dl(1+f_{i}k^{i}%
),\label{Vr_DLaprox}%
\end{equation}
where $dl=(-\eta_{ik}dx^{i}dx^{k})^{1/2}$ is the Euclidean spatial element 
and
$k^{i}=dx^{i}/dl$ is a unit vector of the direction with respect to $dl$.

The phase shift in the interference of two coherent light beams on
a rotating frame was observed by Sagnac \cite{Vr_Post}. For a
relativistic explanation of the effect it is usually postulated
,that space-time in any frames of reference is pseudo - Euclidean
\cite{Vr_Ashtekar}, \cite{Vr_Ananden}. The motion in NIFR is
considered as the relative one in absolute pseudo- Euclidean
space-time.

However, for an $ isolated$ observer (in a ''black box'') in the
rotating frame, who proceeds from the notion of space and time relativity in
the BLMP sense, the observed aniso\-tro\-py in the time of light propagation
(which from his viewpoint contradicts  the experiments of Michelson - Morley
type) is not a trivial effect. It must have some ''internal'' physical 
explanation.

A rigid disk rotating in the plane $xy$ with angular velocity
$\Omega$ around the axis $z$ is approximately identical to the
NIFR described in example 2 of Sec.3. It follows from
eq. (\ref{Vr_DLaprox})  that the spatial element in the NIFR is
anisotropic. We will show that the speed of light in the
noninertial frame of reference is anisotropic too.

A triple of space basis vectors, necessary to compare  the direction of a
given
vector 
from viewpoints of the NIFR and IFR, are not defined above in each point of 
the NIFR. However, it
does not prevent us from comparing the lengths of the vectors. In particular,
 we
can find a dependency between the speeds of a particle in the NIFR and IFR.

The speed of the motion of a particle in the NIFR is
\begin{equation}
v^{\prime}= (-\eta_{\alpha\beta}\overline{u}^{\alpha} \overline{u}^{\beta
})^{1/2} + f_{\alpha}\overline{u}^{\alpha} ,\label{Vr_v'}%
\end{equation}
where $u^{\alpha}= dx^{\alpha}/d\tau$ is 4-velocity of the particle.

The term under the square root is given by
\begin{eqnarray}
-H_{\alpha\beta}u^{\alpha}u^{\beta}
=-\eta_{\alpha\beta}u^{\alpha}u^{\beta
}-\eta_{\mu\nu}\zeta^{\mu}\zeta^{\nu}(\overset{\ast}{\zeta}_{\alpha}v^{\alpha
})^{2}\label{Vr_H1} \nonumber \\
 =-\eta_{\alpha\beta}u^{\alpha}u^{\beta}-c^{2}(d\eta/ds)^{2}+2c^{2}(d\tau
_{0}/d\tau),
\end{eqnarray}
where we have used the following equalities:
\begin{eqnarray}
\overset{\ast}{\zeta}_{\alpha}dx^{\alpha}/d\tau  =c,\\
\zeta_{\alpha}   =\nu_{\alpha}d\eta/ds,\\
\zeta_{\alpha}dx^{\alpha}/d\tau_{0}   =c\ \ \\
\tau_{0}   =c^{-1}u_{\mu}dx^{\mu}%
\end{eqnarray}
The first term in eq. (\ref{Vr_H1}) for a photon is equal to zero
and we obtain with accuracy up to $V/c$ that
$(-\eta_{\alpha\beta}\overline{u}^{\alpha
}\overline{u}^{\beta})^{1/2}=c$.

In the same approximation $f_{\alpha}\overline{u}^{\alpha}=\overline
{f}_{\alpha}u^{\alpha}= c f_{i}k^{i}$.

Thus, in the used coordinates system the speed of the photon in the NIFR 
with
accuracy up to $V/c$ is
\begin{equation}
v^{\prime}_{ph}=c(1+f_{i}k^{i}).\label{Vr_vNIFR}%
\end{equation}

Consider a disk rotating with the constant angular velocity $\Omega$ around
the $z$ axis. Let $r$ and $\theta$ be the coordinates, defined by the
equations
\begin{equation}
x=r\cos(\varphi),y=r\sin(\varphi),\varphi=\theta+\Omega t.\label{Vr_7/-}%
\end{equation}
In the coordinate system $(r,\theta,z,t)$ the space - time metrical
differential form $ds$ in the rotating frame is of the form
\begin{equation}
ds=d\eta-[\Omega r^{2}/(2c)]d\theta-[\Omega^{2}r^{2}/(2c^{2})]dx^{0}%
.\label{Vr_7/10}%
\end{equation}
where $d\eta$ is the pseudo - Euclidean metric form:
\begin{eqnarray}
\label{Vr_7/11}d\eta^{2}   =[1-(\Omega r)^{2}/c^{2}](dx^{0})^{2}-(dr)^{2}%
-r^{2}(d\theta)^{2}\label{Vr_detainsphercoord} \nonumber \\
 -2(r^{2}\Omega/c)d\theta\,dx^{0}-dz^{2}.\;\;
\end{eqnarray}

In this coordinates system eq. (\ref{Vr_uinchroncoord}) is
satisfied with accuracy at least up to $V/c$.  In virtue of
equations (\ref{Vr_DLaprox}) and (\ref{Vr_vNIFR}) the time of the
motion of light from the point $x^{i}$ to $x^{i} + dx^{i}$ is
$dL/v_{p} = c^{-1}dl(1 + 2 f_{i} k^{i} )$. It follows from eq. 
(\ref{Vr_detainsphercoord}) that in the used coordinates system
\[
f_{i}k^{i}= -\frac{\Omega r}{2c} \frac{d\theta}{2\pi}.
\]
For this reason the difference in the time interval between light
propagation around the rotating disk in a clockwise and
counterclockwise direction is $4 \pi r^{2} \Omega/ c^{2} $, which
gives the Sagnac phase shift \cite{Vr_Post}. Thus, the Sagnac
effect for the isolated observer in the rotating frame can be
treated as caused by the Finslerian metric of space-time in
noninertial frames of reference.

\section{ Inertial Forces}

Let us show that the existence of the inertial forces in NIFR can
be interpreted as the exhibition of the Finslerian connection of
space-time in such frames.

According to our initial assumption in Section 3,the differential
equations of motion in an IFR of the point masses, forming the
reference body of the NIFR,are the geodesic lines of space-time in
NIFR. These equations can be found from the variational principle
$\delta\int ds = 0$. The equations are of the form
\begin{equation}
\label{Vr_11/29}d\zeta^{\alpha}/ds + G^{\alpha}(x,\zeta) = 0 ,
\end{equation}
where $\zeta^{\alpha}$ is the 4-velocity of the point mass, the world line 
of
which is $x^{\alpha} = x^{\alpha}(s)$, and
\begin{equation}
\label{Vr_11/-}G^{\alpha}(x,\zeta) = \Gamma^{\alpha}_{\beta\gamma}
\zeta^{\alpha} \zeta^{\gamma} + B^{\alpha}_{\beta} \zeta^{\beta} +
\zeta^{\alpha}\, \rho\, d(\rho^{-1})/ds,
\end{equation}
where
\[
\rho= d\eta/ds = (\eta_{a\beta} \zeta^{\alpha}\zeta^{\beta})^{1/2},\;
B^{\alpha}_{\beta} = \eta^{\alpha\delta} B_{\delta\beta}
\]
and
\[
B_{\delta\beta} = \partial f_{\beta}/ \partial x ^{\delta} -  \partial
f_{\delta} / \partial x^{\beta} .
\]

In the Finslerian space-time a number of connections can be
defined according to eq. (\ref{Vr_11/29}) \cite{Vr_Rund}. In
particular, this equation can be interpreted in the sense that in
the NIFR space-time the absolute derivative of a vector field
$\xi^{a}(x)$ along the world line $x^{\alpha} = x^{\alpha }(s)$ is
of the form
\begin{equation}
\label{Vr_11/30}D \xi^{\alpha}/ds = d\xi^{\alpha} /ds +
G^{\alpha}_{\beta} (x,dx/ds) \xi^{\beta},
\end{equation}
where
\[
G^{\alpha}_{\beta}(x, dx/ds) = B^{\alpha}_{\beta}+ \Gamma^{\alpha}%
_{\beta\gamma} dx^{\gamma}/ds + \rho\, d \rho^{-1}/ds .
\]

Equations (\ref{Vr_11/30}) define a connection of Laugvitz type
\cite{Vr_Rund} in space-time of the NIFR, which is nonlinear
relative to $dx^{\alpha}$. The change in the vector $\xi^{\alpha}$
due to an infinitesimal parallel transport is
\begin{equation}
\label{Vr_11-1}d\xi^{\alpha} = - G^{\alpha}_{\beta}(x,dx)
\xi^{\beta} ,
\end{equation}

Consider the motion of a particle of the mass $m_{p}$, in a NIFR,
unaffected by forces of any kind in the laboratory (inertial)
frame of reference. The differential equations of motion of such a
particle can be found from the variational principle $\delta\int d
\eta= 0$. Since $ds = d\eta- f_{\alpha} dx^{\alpha}$, the
equations of motion are
\begin{equation}
\label{Vr_11/31}Du^{\alpha} /ds = B^{\alpha}_{\beta} u^{\beta} .
\end{equation}

As an example, consider the nonrelativistic disk rotating in the
$xy$ plane about the $z$ axis with the angular velocity $\Omega$.
The equations of motion (\ref{Vr_11/29}) are
\begin{equation}
\label{Vr_11/32}d \vec{\zeta}/dt + \vec{\Omega} \times\vec{r} = 0
,
\end{equation}
where $\vec{r} = \{x,y,z\}$ and the coordinates origin coincides
with the disk center.  The absolute derivative (\ref{Vr_11/30}) of
a vector $\vec{\xi}$ is given by
\begin{equation}
\label{Vr_12/33}D\vec{\xi}/dt = d \vec{\xi} /dt - \vec{\Omega}
\times\vec{ \xi}.
\end{equation}
and the equations of motion (\ref{Vr_11/31}) of the considered
particle in the NIFR are
\begin{equation}
\label{Vr_12/34}D \vec{v} /dt = - \vec{\Omega} \times\vec{ v} ,
\end{equation}
where $\vec{v}= \{ \overset{\cdot}{x}\, \ \overset{\cdot}{y}, 
\ \overset{\cdot}{z} \}$.

Next, for the 4-velocity $u^{\alpha}$ we have
\begin{equation}
\label{Vr_12/35}u^{\alpha} = \overline{u}^{\alpha} +
\zeta^{\alpha} ,
\end{equation}
where $\overline{u}^{\alpha}$  is the spatial velocity of the
particle in the NIFR.  In the nonrelativistic limit
eq. (\ref{Vr_12/35}) can be written in the form
\begin{equation}
\label{Vr_12/36}\vec{v} = \overline{\vec{v}} + \vec{\zeta} ,
\end{equation}
where $\overline{\vec{v}}$ is the "relative" velocity of the
particle and $\vec{\zeta}= \vec{\nu} $ is the velocity of the disk
point in the laboratory frame.  Substituting (\ref{Vr_12/36}) in
(\ref{Vr_12/34}), we find that
\begin{equation}
\label{Vr_12/37}D \overline{\vec{v}}/dt = -
D\overline{\vec{\zeta}} /dt - \vec{\Omega} \times\vec{v} -
\vec{\Omega} \times\vec{\zeta}.
\end{equation}

The value $D\overline{\vec{v}}/dt$ is an acceleration of the considered
particle in the used NIFR found with the help of measuring instruments.  
The
velocities field $\vec{\zeta}$ of the disk points is given by
\begin{equation}
\vec{\zeta} = \vec{\Omega} \times\vec{r}.
\end{equation}

Hence, along the particle path we have $d\vec{\zeta} /dt = \vec{\Omega}
\times\vec{v}$ and
\begin{equation}
\label{Vr_12/38}D \vec{\zeta} /dt = d\vec{\zeta}/dt - \vec{\Omega}
\times \vec{\zeta} = \vec{\Omega} \times\overline{\vec{v}} .
\end{equation}

Thus, finally, we find from (\ref{Vr_12/34})
\begin{equation}
\label{Vr_12/39}m_{p}D\overline{\vec{v}}/dt = -2m_{p}(\vec{\Omega}
\times\overline {\vec{v}}) - m_{p} \vec{\Omega} \times( \vec{\Omega}
\times\vec{r}) .
\end{equation}
We arrived at the nonrelativistic equations of motion of a point
in a rotating frame \cite{Vr_Syng}. The right-hand side of
eq. (\ref{Vr_12/39}) is the ordinary expression for the Coriolis
forces and the centrifugal force in the rotating frames. See also
\cite{Vr_Verozub} ).

Thus, in the nonrelativistic limit the Finslerian space-time in
NIFR manifests itself in the structure of vector derivatives with
respect to the time $t$. It should be noted that
eq. (\ref{Vr_12/33}) is considered sometimes in classical dynamics
nominally \cite{Vr_Syng} just for the derivation of the inertial
forces in the NIFRs.

\section{Relativity of Inertia}


A clock, which is in a NIFR at rest, is unaffected by acceleration in space -
time of the frame. The change in rate of the ideal clock is a real 
consequence
of the difference between the space - time metrics in the IFR and NIFR. It is
given by the factor $\sigma=ds/d\eta$ from the equation $ds=\sigma d\eta$. 
For
the rotating disk of the radius $R$ \ $\sigma=1-\omega^{2}R^{2}/2c^{2}$ which
gives rise to the observed red shift in the well known Pound - Rebka - Snider 
experiments.

We consider here another experimentally verifiable consequence of the above 
theory.

Let $p^{\alpha} = m_{p} c\ dx^{\alpha} / d \eta$ be 4-momentum of a particle 
in the
IFR. Using $3+1$ decomposition of space-time in the NIFR we have
\begin{equation}
p^{\alpha} = \overline{p}^{\alpha} + E \zeta^{\alpha}.
\end{equation}
From the viewpoint of an observer in the NIFR the spatial
projection $\overline{p}^{\alpha}$ should be identified with the
momentum, and the quantity $cE$ with the energy ${\mathcal{E}}$ of
the particle. It is obvious that $E=\overset{\ast}{\zeta}_{\alpha}
p^{\alpha}$.

Therefore, the energy of the particle in the NIFR is
\begin{equation}
{\mathcal{E}} = m_{p} Q c^{2} \overset{\ast}{\zeta}_{\alpha}
u^{\alpha} ,
\end{equation}
where $Q=ds/d \eta= F(x, dx/d \eta)$. For the particle at rest in
the NIFR $u^{\alpha}= \zeta^{\alpha}$ and we obtain
\begin{equation}
{\mathcal{E}} = m_{p}Qc^{2}%
\end{equation}
Thus, the inertial mass $m^{\prime}_{p}$ of the particle in the NIFR is 
given by
\begin{equation}
m^{\prime}_{p}=Qm_{p}
\end{equation}
The quantity $m^{\prime}_{p}$ coincides with the proportionality factor 
between
the momentum and the velocity of the nonrelativistic particle in the NIFR.

Since $Q$ is the function of $x^{\alpha}$, the inertial mass in
the NIFR is not a constant. For example, on the rotating disk we
have
\begin{equation}
m^{\prime}_{p}= m_{p}\, /(1-\Omega^{2}r^{2}/2c^{2}),
\end{equation}
where $\Omega$ is the rotation angular velocity and $r$ is the distance of 
the
body from the disk center.

The difference between the inertial mass $m_{p}^{eq}$ of a body on the 
Earth
equator and the mass $m_{p}^{pol}$ of the same body on the pole is given by
\begin{equation}
(m_{p}^{eq} - m_{p}^{pol})/m_{p}^{pol} = 1.2 \cdot10^{-12}%
\end{equation}
The dependence of the inertial mass of particles on the Earth longitude can
be observed by the M$\overset{..}{\text{o}}$ssbauer  effect. Indeed, the 
change
$\Delta\lambda$ in the wave length $\lambda$ at the Compton scattering on
particles of the masses $m_{p}$ is proportional to $m_{p}^{-1}$. If this 
value is
measured for gamma - quantums with the help of the 
M$\overset{..}{\text{o}}$ssbauer effect
at a fixed scattering angle, then after transporting the measuring device 
from
the longitude $\varphi_{1}$ to the longitude $\varphi_{2}$ we have
\begin{equation}
\frac{ (\Delta\lambda)_{\varphi_{1}}^{-1} - 
( \Delta\lambda)_{\varphi_{2} } ^{-1}
} { (\Delta\lambda)_{\varphi_{1}}^{-1} } = \Theta \ [ cos^{2}(\varphi_{1}) - 
cos^{2}(\varphi_{2})],
\end{equation}
where $\Theta =1.2 \cdot 10^{-12}$.

\section{Gravitation in Inertial and Proper Reference Frames}

Consider a frame of reference whose reference body is formed by
identical material points $m_{p}$ moving under the effect of the
field $\psi _{\alpha\beta}$. These frames will be called the
proper frames of reference (PRF) of the given field. Any observer,
located in the PFR at rest, moves in space-time of this frame
along the geodesic line of his space-time. This implies that the
space-time metric differential form in the NIFR is given by
eq. (\ref{Vr_5/4})
where $S$ is the action describing describing
 the motion of
particles forming the reference body of the NIFR.

Now suppose \cite{Vr_Thirring} that in pseudo-
Euclidean space-time   gravitation can be described
as a tensor field $\psi_{\alpha\beta(x)}$, and the Lagrangian 
describing motion
of a test particle with the mass $m_{p}$ is of the form
\begin{equation}
L=-m_{p} c [g_{\alpha\beta}(\psi)\;\overset{\cdot}{x}^{\alpha}\;
\overset{\cdot}{x}^{\beta}]^{1/2},
\label{Vr_LagrangianThirr}%
\end{equation}
where $\overset{\cdot}{x}^{\alpha}=dx^{\alpha}/dt$ and $g_{\alpha\beta}$ 
is the symmetric
tensor whose components are the function of $\psi_{\alpha\beta}$.

According to (\ref{Vr_5/4}) the space-time metric differential
form in the PFR is given by
\begin{equation}
ds^{2}=g_{\alpha\beta}(\psi)\;dx^{\alpha}\;dx^{\beta}%
\end{equation}

Thus, the space-time in the PFR is a Riemannian with the curvature other than
zero. Viewed by an observer in the IFR, the motion of the test particle
forming the reference body of the PFR is affected by the force field
$\psi_{\alpha\beta}$. But the observer located in the PRF will not observe 
the
force properties of the field $\psi_{\alpha\beta}$ since he moves in
space-time of the PRF along the geodesic line. For him the presence of the
field $\psi_{\alpha\beta}$ will be displayed in another way --- as 
space-time
curvature differing from zero in these frames, e.g. as a deviation of the
world lines of the neighbouring points of the reference body.

For example, when studying the Earth's gravity, a frame of reference
fixed to the Earth can be considered as an inertial frame if the
forces of inertia are ignored. An observer located in this frame
can consider motion of the particles forming the PRF reference
body in flat space-time on the basis of
eq. (\ref{Vr_LagrangianThirr}) without running into contradiction
with experiments. However, the observer in the PFR (in a comoving
frame for free falling particles) does not find the Earth
gravity as some force field. If he proceeds from the relativity of
space-time, he believes that point particles, forming the reference
body of his reference frame, are the point of his physical space.
They do not affect a force field and their accelerations in his
space-time are equal to zero. In spite of that, he observes a
change in the relative distances of these particles. Such an
experimental fact has apparently the only explanation as
non-relativistic display of the deviations of the geodesic lines
caused by space-time curvature. So, we observe an important fact
that only in proper frames of reference we have an evidence for
gravitation identification with space-time curvature.

Thus, we arrive at the following hypothesis. In inertial frames of
reference, where space-time is pseudo - Euclidean, gravitation is a
field $\psi _{\alpha\beta}$. In the proper frames of reference of
the field $\psi _{\alpha\beta}$, where space-time is Riemannian,
gravitation manifests itself as curvature of space-time and must be
described completely by the geometrical properties of the letter.

If this possibility really takes place in nature, then it will remove an
isolation of the geometrical gravitational theories from the theories of 
other fields.

Of course, eq. (\ref{Vr_5/4}) refers to any classical field. For
instance, space-time in the PRF of an electromagnetic field is
Finslerian. However, since $ds$
depends on the mass $m_{p}$ and charge $e$ of the point masses forming
the reference body, this fact is not of great significance.

It should be noted that the geometrical theory of gravitation in
the PFR is not identical to Einstein's theory. Gravitational
equations should be some  kind of differential equations for the
function $\psi_{\alpha\beta}$ or $g_{\alpha\beta}\;(\psi)$, which
are invariant under a certain set of gauge transformations of
the potentials $\psi_{\alpha\beta}$. Since $g_{\alpha\beta} =
g_{\alpha\beta}\;(\psi)$, the Einstein equations are the equations
both for $g_{\alpha\beta}$ and for $\psi_{\alpha\beta}$. Under the 
transformation $\psi_{\alpha\beta}
\to\overline{\psi}_{\alpha\beta}$ the quantities
$g_{\alpha\beta}\;(\psi)$ undergo some transformations too, and,
as a consequence, the equations of the test particle motion
resulting from eq. (\ref{Vr_LagrangianThirr}) and the Einstein's
equations do not remain invariant.

The equations of motion resulting from
eq. (\ref{Vr_LagrangianThirr}) are at the same time the equations
of a geodesic line of the Riemannian space-time $V_{n}$ of the
dimensionality $n$ with the metric tensor $g_{\alpha\beta
}\;(\psi)$. That is why if the given gauge transformation
$\psi_{\alpha\beta }\rightarrow\overline{\psi}_{\alpha\beta}$
leaves the equations of motion
invariant, then the corresponding transformation $g_{\alpha\beta}%
\rightarrow\overline{g}_{\alpha\beta}$ is a mapping
$V\rightarrow\overline{V}$ of the Riemannian spaces leaving
geodesic lines invariant, i.e. it is a geodesic, (projective)
mapping. Let us assume that not only eq. (\ref{Vr_LagrangianThirr})
but also the field equations contain $\psi_{\alpha\beta}$ only
in the form $g_{\alpha\beta}\;(\psi)$, then it becomes clear that
the gauge-invariance of the equations of motion will be ensured if
the field equations are invariant with respect to geodesic
mappings of the Riemannian space $V_{n}$. Thus, if we start from
eq. (\ref{Vr_LagrangianThirr}), then the gravitational field
equations as well as the physical field characteristics must be
invariant with respect to geodesic (projective) mappings of the
Riemannian space-time $V_{n}$ with the metric tensor
$g_{\alpha\beta}\;(\psi)$.

Simplest equations of that kind are analyzed in the next
Section.

\section{The geodesic-invariant equations of gravitation}

In accordance with the basic principles the space-time in the PRF of the
gravitational field $\psi_{\alpha\beta}$ is the Riemannian $V_{n}$ of
dimension n=4 whose metric tensor is defined up to geodesic (projective)
mappings $g_{\alpha\beta}\rightarrow\overline{g}_{\alpha\beta}$, generated by
the gauge transformations $\psi_{\alpha\beta}\rightarrow
\overline{\psi}_{\alpha\beta}$.
The geodesic transformations of the metric tensor
$g_{\alpha\beta}$ are given by Levi-Chivita equations
\cite{Vr_Levi-Chivita}, \cite{Vr_Sinukov}:
\begin{equation}
\nabla_{\alpha}\overline{g}_{\beta\gamma}=2\varphi_{\alpha}\overline{g}%
_{\beta\gamma}+\varphi_{\beta}\overline{g}_{\gamma\alpha}+\varphi_{\gamma
}\overline{g}_{\alpha\beta}\;, \label{Vr_levichiv}%
\end{equation}
where
\[
\varphi_{\alpha}=\frac{1}{2(n+1)}\frac{\partial}{\partial x^{\alpha}}%
\lg\left|  \frac{\overline{g}}{g}\right|  \;,
\]%
\[
\nabla_{\alpha}\overline{g}_{\beta\gamma}=\partial\overline{g}_{\beta\gamma
}/\partial x^{\alpha}-\Gamma_{\alpha\beta}^{\mu}\overline{g}_{\mu\gamma
}-\Gamma_{\alpha\gamma}^{\mu}\overline{g}_{\mu\beta}\;,
\]
$g=\det|g_{\alpha\beta}|$

The Christoffel symbols and the curvature tensor also do not
remain invariant. In particular, the Christoffel symbols are
transformed as follows \cite{Vr_Sinukov}:
\begin{equation}
\overline{\Gamma}_{\alpha\beta}^{\gamma}=\Gamma_{\alpha\beta}^{\gamma}%
+\delta_{\alpha}^{\gamma}\phi_{\beta}+\delta_{\beta}^{\gamma}\phi_{\alpha
};\label{Vr_GammaTransformUnderProjeciveMappings}%
\end{equation}
where $\phi_{\alpha}(x)$ is a vector field.

However, some objects, which are invariant under geodesic (projective)
mappings of the space $V_{n}$, also can be defined. Just these 
gauge-invariant
objects have a physical sense in the theory under consideration.

The simplest gauge-invariant object is the Thomas symbols
\cite{Vr_Berwald}:

\begin{equation}
\Pi _{\alpha \beta }^{\gamma } = \Gamma _{\alpha \beta }^{\gamma } -
(n+1)^{-1} \left [ \delta _{\alpha }^{\gamma } \Gamma _{\beta \mu }^{\mu } +
\delta_{\beta }^{\gamma } \Gamma _{\alpha \mu }^{\mu } \right ]\;.
\end{equation}

The Thomas symbols are transformed as a tensor only with respect to the
projective coordinate transformations. However, by replacing the ordinary
derivatives to the covariant ones for the metric $d\sigma^{2}$, a tensor
$B_{\alpha\beta}^{\gamma}$ can be received. This object also can be written 
as
follows
\begin{equation}
B_{\alpha\beta}^{\gamma}=\Pi_{\alpha\beta}^{\gamma}-\overset{\circ}{\Pi
}_{\alpha\beta^{\gamma}}\;, \label{Vr_tens B}%
\end{equation}
where $\overset{\circ}{\Pi}_{\alpha\beta}^{\gamma}$ are Thomas symbols in 
the
$E_{n}$:%

\begin{equation}
\stackrel{\circ }{\Pi }_{\alpha \beta }^{\gamma }=\stackrel{\circ }{\Gamma }%
_{\alpha \beta }^{\gamma }-(n+1)^{-1}\left[ \delta _{\alpha }^{\gamma }%
\stackrel{\circ }{\Gamma }_{\epsilon \beta }^{\epsilon }+\delta _{\beta
}^{\gamma }\stackrel{\circ }{\Gamma }_{\epsilon \alpha }^{\epsilon }\right] ,
\end{equation}

This geodesic - invariant tensor will be named the strength tensor of a
gravitational field. Note that the equality $B_{\alpha\gamma}^{\gamma}=0$ is
satisfied identically.

According to eq. (\ref{Vr_LagrangianThirr}) the differential
equations of motion of the test particle are given by
\begin{equation}
\frac{d^{2}x^{\gamma}}{ds^{2}}+\Gamma_{\alpha\beta}^{\gamma}
\frac{dx^{\alpha}%
}{ds}\frac{dx^{\beta}}{ds}=0\;, \label{Vr_geodeseq}%
\end{equation}
where $\Gamma_{\alpha\beta}^{\gamma}$ are the Christoffel symbols
in $V_{n}$. Following \cite{Vr_Berwald} we will show how one can
define a geodesic - invariant connection and curvature in the
spaces under consideration.

Let us define a scalar parameter $p$ on the geodesic lines, that remains
unaltered by geodesic mapping of $V_{n},$ by means of a differential 
equation%

\begin{equation}
\{p,s\}=-2 \epsilon\Gamma_{\alpha\beta}^{0}\frac{dx^{\alpha}}{ds}%
\frac{dx^{\beta}}{ds}, \label{Vr_DefinitionGammaZero}%
\end{equation}
where
\begin{equation}
\{p,s\}=\frac{\frac{d^{2}p}{ds^{2}}}{\frac{dp}{ds}}-\frac{3}{2}\left(
\frac{\frac{d^{2}p}{ds^{2}}}{\frac{dp}{ds}}\right)  ^{2},
\end{equation}
$\epsilon$ is a nonzero constant  and $\Gamma_{\alpha\beta}^{0}$ are a
given
function of $x^{\alpha}$ symmetric in low indices:
\begin{equation}
\Gamma_{\alpha\beta}^{0}=\Gamma_{\beta\alpha}^{0}.
\end{equation}
By eq. (\ref{Vr_DefinitionGammaZero}) the parameter $p$ is defined
as the function of $s$ up to linear fractional transformations.

Since the parameter $p$ must be a scalar, the object $\Gamma_{\alpha\beta}%
^{0}$ is the components of the covariant tensor of rank 2.

Let $\overline{\Gamma}_{\alpha\beta}^{0}$ and $d\overline{s}$ be the
components of $\Gamma_{\alpha\beta}^{0}$ and the line element $ds$,
respectively, after some geodesic mapping of space-time $V_{n}$. Then the new
geodesic equations are given by
\begin{equation}
\frac{d^{2}x^{\gamma}}{d\overline{s}^{2}}+\overline{\Gamma}_{\alpha\beta
}^{\gamma}\frac{dx^{\alpha}}{d\overline{s}}\frac{dx^{\beta}}{d\overline{s}%
}=0\;.\label{Vr_geodesicEquationAfterGeodesicMappings}%
\end{equation}
On the other hand, after a geodesic mapping eqs.(\ref{Vr_geodeseq})
are
\begin{equation}
\frac{d^{2}x^{\gamma}}{d\overline{s}^{2}}+\frac{\overline{s}^{\prime\prime}%
}{(\overline{s}^{\prime})^{2}}\frac{dx^{\alpha}}{d\overline{s}}+\Gamma
_{\alpha\beta}^{\gamma}\frac{dx^{\alpha}}{d\overline{s}}\frac{dx^{\beta}%
}{d\overline{s}}=0,\nonumber
\end{equation}
where $\overline{s}^{\prime}=d\overline{s}/ds$ ,
$\overline{s}^{\prime\prime }=d^{2}\overline{s}/ds^{2}$, and these
equations must be identical to
(\ref{Vr_geodesicEquationAfterGeodesicMappings}). With regards to
(\ref{Vr_GammaTransformUnderProjeciveMappings}), this yields
\begin{equation}
d\overline{s}/ds=\exp \left(-2\int\phi_{\alpha}dx^{\alpha}
\right)\label{Vr_TransformOfS-parameter}%
\end{equation}

Then, by putting
\begin{equation}
\{p,\overline{s}\}=-2\epsilon\overline{\Gamma}_{\alpha\beta}^{0}%
\frac{dx^{\alpha}}{\overline{ds}}\frac{dx^{\beta}}{\overline{ds}}%
\end{equation}
and by taking notice of
\begin{equation}
\{p,\overline{s}\}=(ds/\overline{ds})^{2}\left[  \{p,s\}-\{\overline
{s},s\}\right]  ,
\end{equation}
we find (since $p$ must be the invariant under geodesic mappings) the
equations of transformation of $\Gamma_{\alpha\beta}^{0}$ under geodesic
mappings:
\begin{eqnarray}
\overline{\Gamma}_{\beta\gamma}^{0}   =\Gamma_{\beta\gamma}^{0}-\epsilon
^{-1}\left\{  \frac{1}{2}\left(  \frac{\partial\phi_{\beta}}{\partial
x^{\gamma}}+\frac{\partial\phi_{\gamma}}{\partial x^{\beta}}\right)  \right\}
\nonumber\\
 -\phi_{\kappa}\Gamma_{\beta\gamma}^{\kappa}+\phi_{\beta}\phi_{\gamma}%
\label{transgam0}
\end{eqnarray}

Now on every geodesic we define a \emph{gauge variable} (''5 th
coordinate'') by substitution
\begin{equation}
x^{4}=-\frac{1}{2\epsilon}\log\frac{ds}{dp}. \label{Vr_GaugeVariable}%
\end{equation}

Let $\overline{s}$ be another parameter on the geodesic and
\begin{equation}
\overline{x}^{4}=-\frac{1}{2\epsilon}\log\frac{d\overline{s}}{dp}.
\end{equation}
is the gauge variable corresponding to the transition
$p\rightarrow \overline{s}$. Then it follows from
(\ref{Vr_TransformOfS-parameter}) that for any path
\begin{equation}
\overline{x}^{0}=x^{0}+\epsilon^{-1}\int_{q}\phi_{\alpha}dx^{\alpha}+C,
\end{equation}
where integration is performed from an arbitrary fixed point $q$ on the 
geodesic
along the curve and $C$ is the arbitrary constant.

Using the relation
\begin{equation}
\{v,u\}=-\left(  \frac{dv}{du}\right)  ^{2}\{u,v\},
\end{equation}
we have
\begin{equation}
\{s,p\}=-2\epsilon\Gamma_{\alpha\beta}^{0}\frac{dx^{\alpha}}{dp}%
\frac{dx^{\beta}}{dp}. \label{Vr_SP}%
\end{equation}
On replacing in this equation and the geodesic equations
(\ref{Vr_geodeseq}) the derivations $ds/dp$, $d^{2}s/dp^{2},$
$d^{3}s/dp^{3}$ by $x^{0},$ $dx^{0}/dp$, $d^{2}x^{0}/dp^{2}$
according to the definition of the gauge variable
(\ref{Vr_GaugeVariable}) we find that the geodesic equation
(\ref{Vr_geodeseq}) and the equation (\ref{Vr_SP}) which yields
the definition of the parameter $p$ can be written as
\begin{equation}
\frac{d^{2}x^{\alpha}}{dp^{2}}+\Gamma_{\beta\gamma}^{\alpha}\frac{dx^{\beta}%
}{dp}\frac{dx^{\gamma}}{dp}+2\epsilon\frac{dx^{4}}{dp}\frac{dx^{\alpha}}%
{dp}=0\;, \label{Vr_geodeeq2}%
\end{equation}%
\begin{equation}
\frac{d^{2}x^{4}}{dp^{2}}+\epsilon\left(  \frac{dx^{4}}{dp}\right)
^{4}+\Gamma_{\beta\gamma}^{4}\frac{dx^{\beta}}{dp}\frac{dx^{\gamma}}{dp}=0\;,
\label{Vr_defproecpar}%
\end{equation}
respectively. These equations can be united as an equation of a geodesic in
5-dimensional space-time
\begin{equation}
\frac{d^{2}x^{A}}{d\pi^{2}}+\Gamma_{BC}^{A}\frac{dx^{B}}{d\pi}\frac{dx^{C}%
}{d\pi}=0 \label{Vr_5eqgeodesic}%
\end{equation}

The capital indices run from $0$ to $4$ and $\Gamma_{B0}^{A}=\Gamma_{0B}%
^{A}=M\delta_{B}^{A}$ where $\delta_{B}^{A}$ is the Kronecker
symbols.

Suppose the functions $\phi_{\alpha}$ satisfy the conditions:
\begin{equation}
\partial\phi_{\alpha}/\partial x^{\beta}-\partial\phi_{\beta}/\partial
x^{\alpha}=0.
\end{equation}

If we consider the transformations
\begin{equation}
\overline{x}^{\alpha}=f^{\alpha}\;(x^{0},x^{1},x^{2},x^{3})
\label{Vr_coordtransform}%
\end{equation}
and
\begin{equation}
\overline{x}^{0}=x^{0}+\epsilon^{-1}\int_{q}\phi_{\alpha}dx^{\alpha}+C,
\label{Vr_x4 transform}%
\end{equation}
where $\phi_{\alpha}dx^{\alpha}$ is the exact differential of a
function of $x^{\alpha}$, as admissible coordinate transformations
in the $n+1$ dimensional manifold ${\mathcal{M}}_{5}$, then 
eqs. (\ref{Vr_5eqgeodesic}) can be regarded as the differential
equations of the geodesic lines in homogeneous coordinates of
projective geometry. A theory of the projective connection in such
a way has been considered by \cite{Vr_Berwald}$\div$
\cite{Vr_Weblen} and other authors in 1921-1937.

The object components $\Gamma_{BC}^{A}$ are transformed as follows:%

\begin{equation}
\overline{\Gamma}_{BC}^{A}=\left(  \Gamma_{ED}^{F}\frac{\partial x^{E}%
}{\partial\overline{x}^{B}}\frac{\partial x^{D}}{\partial\overline{x}^{C}%
}+\frac{\partial^{2}x^{F}}{\partial\overline{x}^{B}\partial\overline{x}^{C}%
}\right)  \frac{\partial\overline{x}^{A}}{\partial x^{F}}%
\label{Vr_transform5conn}%
\end{equation}

The components of $\Gamma_{BC}^{A}$ are coefficients of the
projective connection \cite{Vr_Berwald}. The object
\begin{eqnarray}
{\mathcal{R}}_{BCD}^{A}=\partial\Gamma_{BC}^{A}/\partial
x^{D}-\partial \Gamma_{BD}^{A}/\partial x^{C} \nonumber \\
+\Gamma_{MD}^{A}\Gamma_{BC}^{M}-  
\Gamma_{MC}^{A}\Gamma_{BD}^{M}
\label{Vr_curvature}%
\end{eqnarray}
is transformed as the tensor relative to the coordinate
transformations in ${\mathcal{M}}_{5}$. The components of the
tensor ${\mathcal{R}}_{BCD}^{A}$ vanish identically save
${\mathcal{R}}_{\alpha\beta\gamma}^{4}$ and
\begin{eqnarray}
{\mathcal{R}}_{\beta\gamma\delta}^{\alpha}=R_{\beta\gamma\delta}^{\alpha
}+\epsilon\Gamma_{\beta\gamma}^{4}\delta_{\delta}^{\alpha}-
\epsilon
\Gamma_{\beta\delta}^{4}\delta_{\gamma}^{\alpha}%
,\label{Vr_ProjectiveCurvatureTensor}%
\end{eqnarray}
where $R_{\beta\gamma\delta}^{\alpha}$ is the curvature tensor of the affine
connection in $V_{4}$. It has the following properties
\begin{eqnarray}
{\mathcal{R}}_{\alpha\gamma\delta}^{\alpha}  =0,  \\
{\mathcal{R}}_{\beta\gamma\delta}^{\alpha}+{\mathcal{R}}_{\beta\delta\gamma
}^{\alpha}   =0,\\
{\mathcal{R}}_{\beta\gamma\delta}^{\alpha}+{\mathcal{R}}_{\gamma\delta\beta
}^{\alpha}+{\mathcal{R}}_{\delta\beta\gamma}^{\alpha}   =0.
\end{eqnarray}

It follows from eq. (\ref{Vr_ProjectiveCurvatureTensor}) that the
contracted tensor is given by
\begin{equation}
{\mathcal{R}}_{\beta\gamma}=R_{\beta\gamma}+\epsilon(n-1)\Gamma_{\beta\gamma
}^{0}.
\end{equation}
It does not change with geodesic mappings of $V_{4}$.

The equations
\begin{equation}
{\mathcal{R}}_{\beta\gamma}=0
\label{Vr_GeodesicInvariantGeneralisationEinstenEqInVacuum}%
\end{equation}
are the simplest geodesic - invariant generalization of the Einstein 
vacuum equations.

Depending on a choice of the object $\Gamma_{\beta\gamma}^{0}$ and 
$\epsilon$
we obtain a specific variant of the theory. In this paper we will assume 
that
$\epsilon= 1$ and
\begin{gather}\nonumber
\Gamma_{\alpha\beta}^{0}=-\frac{1}{(n+1)}\times \;\;\; \\
\left[  \frac{1}{2}\left(  \partial
Q_{\alpha}/\partial x^{\beta}+\partial Q_{\beta}/\partial x^{\alpha}\right)
-\Gamma_{\alpha\beta}^{\gamma}Q_{\gamma}-Q_{\alpha}Q_{\beta}
\right], \; \; 
\label{Vr_Gamm00}%
\end{gather}
where $Q_{\alpha}=$
$\Gamma_{\beta\alpha}^{\beta}-\overset{\circ}{\Gamma
}_{\beta\alpha}^{\beta}.$ The object defined in this way has the required
properties under geodesic mappings (\ref{transgam0}).
Equation (\ref{Vr_GeodesicInvariantGeneralisationEinstenEqInVacuum}) can
be written in the form
\begin{equation}
B_{\beta\gamma;\alpha}^{\alpha}-B_{\beta\mu}^{\nu}\;B_{\gamma\nu}^{\mu}=0\;,
\label{Vr_MyEqVacuum}%
\end{equation}
where the semi-colon denotes a covariant derivative in the
Pseudo - Euclidean space-time $E_{4}$. These equations were proposed first
in \cite{Vr_Verozub91} from another viewpoint.

Equations (\ref{Vr_MyEqVacuum}) are the system of the
differential equations for the geodesic- invariant tensor
$B_{\beta\gamma}^{\alpha}$ (or for the functions $g_{\alpha\beta}$)
which are defined up to arbitrary geodesic mappings. The
coordinate system is defined by the used measurement instruments
and is given. The equations do not contain the functions
$\psi_{\alpha\beta}$ explicitly.

The simplest way of obtaining equations for $\psi_{\alpha\beta}$ is to set%

\begin{equation}
B_{\beta\gamma}^{\alpha}=\nabla^{\alpha}\psi_{\beta\gamma}-(n+1)^{-1}\left(
\delta_{\beta}^{\alpha}\nabla^{\sigma}\psi_{\sigma\gamma}+\delta_{\gamma
}^{\alpha}\nabla^{\sigma}\psi_{\beta\sigma}\right)  , \label{Vr_BbyPsi}%
\end{equation}
where $\nabla^{\alpha}$ is the covariant derivative in flat
space-time. The identity
$B_{\alpha\gamma}^{\gamma}\equiv0$ is satisfied as it is to be expected
according 
to the definition of the tensor $B_{\alpha\gamma}^{\gamma}$. 

Then, at the gauge condition
$\nabla^{\sigma}\psi_{\sigma\gamma}=0$ eq. 
(\ref{Vr_MyEqVacuum}) are given by

\begin{eqnarray}
\square\psi_{\alpha\beta}-\nabla^{\sigma}\psi_{\alpha\gamma}\,\nabla^{\gamma
}\psi_{\sigma\beta}   =0\label{Vr_EqsPsiVac}\\
\nabla^{\sigma}\psi_{\sigma\gamma}   =0,\nonumber
\end{eqnarray}
where $\square$ is the covariant  Dalamber operator in pseudo - Euclidean 
space-time.

It is natural to suppose that with the presence of matter these 
equations are given by
\begin{eqnarray}
\square\psi_{\alpha\beta}
=\varkappa(T_{\alpha\beta}+t_{\alpha\beta }) ,
\label{Vr_EqPsiMatter}\\ \nabla^{\sigma}\psi_{\sigma\gamma} =0,\nonumber
\end{eqnarray}
where $\varkappa=8\pi G/c^{4},$ 
$t_{\alpha\beta}=\varkappa^{-1}\nabla^{\sigma
}\psi_{\alpha\gamma}\,\nabla^{\gamma}\psi_{\sigma\beta}$ and 
$T_{\alpha\beta}$
is the matter tensor of the energy-momentum.

Obviously, the equality
\begin{equation}
\nabla^{\beta}(T_{\alpha\beta}+t_{\alpha\beta})=0 \label{Vr_ConservLawPsi}%
\end{equation}
is valid. Therefore, the magnitude $t_{\alpha\beta}$ can be interpreted as 
the
energy-momentum tensor of a gravitational field. 
\footnote{It should be noted that, when we introduce it in some way, 
we cannot be sure apriori that the
equation for $\psi_{\alpha\beta}$ yields all solutions of the equations for
$B_{\beta\gamma}^{\alpha}$. We may introduce a
potential $\psi_{\alpha\beta}$ also in another way.  }
\sloppy
\section{Spherically-Symmetric Gravitational Field.}

Let us find the spherically symmetric solution of
eqs.(\ref{Vr_MyEqVacuum}) in an inertial frame of reference,
where space-time is supposed to be  pseudo - Euclidean.

Because of the gauge (geodesic) invariance, additional conditions
can be imposed on the tensor $g_{\alpha\beta}$. In particular
\cite{Vr_Verozub91}, under the conditions
\begin{equation}
Q_{\alpha}=\Gamma_{\alpha\sigma}^{\sigma}-\overset{\circ}{\Gamma}%
_{\alpha\sigma}^{\sigma}=0 \label{Vr_gaugeConditions}%
\end{equation}
eqs. (\ref{Vr_MyEqVacuum}) will be reduced to the Einstein vacuum 
equations $R_{\alpha\beta}=0$, where $R_{\alpha\beta}$ is the
Ricci tensor. Let us choose a spherical coordinate system. Then, if the
test particle Lagrangian is invariant under the mapping
$t\rightarrow-t$, the fundamental metric form of space-time
$V_{4}$ can be written as
\begin{eqnarray}
ds^{2}=-A(dr)^{2}-B[(d\theta)^{2}+\sin^{2}(\theta)(d\varphi)^{2}]\nonumber\\
+C(dx^{0})^{2}, \label{Vr_ds}%
\end{eqnarray}
where $A$, $B$ and $C$ are the functions of the radial coordinate $r$.
Proceeding from the above- stated, we shall find the functions $A$, $B$ and
$C$ as the solution of the equations system
\begin{equation}
R_{\alpha\beta}=0 \label{Vr_Einstein Eqs}%
\end{equation}
and
\begin{equation}
Q_{\alpha}=0, \label{Vr_AditionalCondition}%
\end{equation}
which satisfy the conditions:
\begin{equation}
\lim\limits_{r\rightarrow\infty}A=1,\;\lim\limits_{r\rightarrow\infty}%
(B/r^{2})=1,\;\lim\limits_{r\rightarrow\infty}C=1. 
\label{Vr_limitConditions}%
\end{equation}

The equations $R_{11}=0$ and $R_{00}=0$ can be written
\cite{Vr_Petrov} as
\begin{equation}
BL_{2}-2CL_{1}=0 \label{Vr_EinstEq1}%
\end{equation}
and
\begin{equation}
B^{\prime}C^{\prime}-2BL_{2}=0, \label{Vr_EinstEq2}%
\end{equation}
where
\[
L_{1}=R_{1212}=B^{\prime\prime}/2-(B^{\prime})^{2}/(4B)-B^{\prime}A^{\prime
}/(4A)\;,
\]

\[
L_{2}=R_{1010}=-C^{\prime\prime}/2+C^{\prime}(AC)^{\prime}/(4AC)
\]
and the differentiation of $A$, $B$ and $C$ with respect to $r$ is denoted by
the primes.

The nonzero eqs.(\ref{Vr_AditionalCondition}) yield
\begin{equation}
B^{2}AC=r^{4}. \label{Vr_BACequal}%
\end{equation}
First, the combination of eqs. (\ref{Vr_EinstEq1}) and
(\ref{Vr_EinstEq2}) yields
\[
4CL_{1}-B^{\prime}C^{\prime}=0,
\]
i.e.
\begin{equation}
2B^{\prime\prime}-(B^{\prime})^{2}/B-B^{\prime}(AC)^{\prime}/(AC)=0. 
\label{Vr_1}%
\end{equation}
Also, taking the logarithm of eq. (\ref{Vr_BACequal}) we obtain
\begin{equation}
(AC)^{\prime}/(AC)=4/r-2B^{\prime}/B=0. \label{Vr_2}%
\end{equation}
Equation (\ref{Vr_1}) then becomes
\[
2B^{\prime\prime}+(B^{\prime})^{2}/B-4B^{\prime}/r=0\;,
\]
or
\begin{equation}
u^{\prime}/u=4/r, \label{Vr_3}%
\end{equation}
where $u=(B^{\prime})^{2}B$. By using (\ref{Vr_limitConditions})
we find
\begin{equation}
B=(r^{3}+{\mathcal{K}}^{3})^{2/3}, \label{Vr_Bequal}%
\end{equation}
where ${\mathcal{K}}$ is a constant.

Next, from eq. (\ref{Vr_EinstEq2}) we find by using
eqs.(\ref{Vr_limitConditions}),
\begin{equation}
C=1-{\mathcal{Q}}/B^{1/2}, \label{Vr_Cequal}%
\end{equation}
where $\mathcal{Q}$ is a constant.

Finally, we can find the function $A$ from eq. 
(\ref{Vr_BACequal}).

Thus, in the spherical symmetric coordinate system the
following functions $A$, $B$ and $C$ are obtained:
\begin{equation}
A=(f^{\prime})^{2}(1-{\mathcal{Q}}/f)^{-1},\;B=f^{2},\;C=1-{\mathcal{Q}}/f,
\label{Vr_ABCequal}%
\end{equation}
where
\[
f=(r^{3}+{\mathcal{K}}^{3})^{1/3}%
\]
and $f^{\prime}=df/dr$.

The nonzero components of the tensor $B_{\alpha\beta}^{\gamma}$ are given by%

\begin{eqnarray}
B_{rr}^{r}   =\frac{1}{2}\frac{A^{\prime}}{A};\quad B_{\theta\theta}%
^{r}=\frac{1}{2}\frac{2rA-B^{\prime}}{A};\nonumber\\
\quad B_{\phi\phi}^{r}   =-\frac{1}{2}\frac{2rA-B^{\prime}%
}{A}\sin^{2}\theta       \nonumber \\
B_{tt}^{r}   =\frac{1}{2}\frac{C^{\prime}}{A};\quad B_{tr}^{t}=\frac{1}%
{2}\frac{C^{\prime}}{C};\\
\quad B_{\theta r}^{\theta}   =\frac{1}{2}\frac{2B-B^{\prime}r}{rB};\quad
B_{\phi r}^{\phi}=\frac{2B-B^{\prime}r}{rB}\nonumber
\end{eqnarray}

If ${\mathcal{K}}=0,$ then $f=r,$ and the solution
(\ref{Vr_ABCequal}) coincides with the Droste-Weyl 
 solution of the
Einstein equations which is known as the Schwarzshild one 
\footnote{A discussion of the difference in these solutions are 
given in \cite{Abrams1}}
.If
$\mathcal{K}=\mathcal{Q},$ it coincides with the originally
Schwarzshild solution \cite{Vr_Schwarzschild}. However, it is
important to understand that from the point of view of the
considered theory, solution (\ref{Vr_ABCequal}) is obtained
in a given (spherical) coordinate system, defined in
pseudo - Euclidean space-time, and that different values of the
constants $\mathcal{Q}$ and $\mathcal{K}$ yield different solutions
of equation (\ref{Vr_MyEqVacuum}) in the same coordinate
system.

The equations of the motion of a test particle resulting from Lagrangian
(\ref{Vr_LagrangianThirr}) is given by%

\begin{equation}
\overset{..}{x}^{\alpha}+(\Gamma_{\beta\gamma}^{\alpha}-c^{-1}\Gamma
_{\beta\gamma}^{0}\overset{.}{x}^{\alpha})\overset{.}{x}^{\beta}\overset{.}%
{x}^{\gamma}=0. \label{Vr_GeodesicLineWithT}%
\end{equation}
In the nonrelativistic limit
$\overset{..}{x}^{r}=-c^{2}\,\Gamma_{00}^{r},$ where
$\Gamma_{00}^{r}=C^{\prime}/2A=r^{4}C^{\prime}/f^{4}C.$ Therefore,
to obtain the Newton gravity law it should be supposed  that at large
$r$ the function $f\approx r$ and
${\mathcal{Q}}=r_{g}=2G\,M/c^{2}$ is the classical Schwarzshild
radius.

At a given constant ${\mathcal{Q}}$ allowable solutions
are obtained by change of the arbitrary constant $\mathcal{K.}$ 
In particular, if we
setting $\mathcal{K} = 0 $ the fundamental form of space-time $V_{4}$ 
coincides
with the Droste-Weyl solution of the Einstein equation \cite{Abrams1} 
(it is
commonly named the Schwarzshild solution) which have an events horizon at
$r=r_{g}$:%

\begin{align}
ds^{2}  & =-\;\frac{dr^{2}}{(1-\frac{r_{g}}{r})}-r^{2}[d\theta^{2}+\sin
^{2}\theta\;d\varphi^{2}]\nonumber\\
& +(1-\frac{r_{g}}{r})\;dx^{0}{}^{2}%
\end{align}
If we setting $\mathcal{K}={\mathcal{Q}}$, the solution coincides with the
original Schwarzshild solution \cite{Vr_Schwarzschild} which have no the 
event
horizon and singularity in the center:

\begin{align}
ds^{2}  & =-\;\frac{f^{\prime2}dr^{2}}{(1-\frac{r_{g}}{f})}-f^{2}[d\theta
^{2}+\sin^{2}\theta\;d\varphi^{2}]\nonumber\\
& +(1-\frac{r_{g}}{f})\;dx^{0}{}^{2},
\end{align}
where $f=($ $r_{g}^{3}$ $+$ $r^{3})^{1/3}$. 

These solutions are related to 
the same coordinate system and are different solutions of the gravitation
equations under consideration.

Of course, a reader can say: however, we can obtain the Droste-Weyl solution
from the original Schwarzshild one by a coordinate transformation.
Supposing it is so. (There is an alternative point of view
\cite{Adams1}). However, in this case we must transform also space and time
intervals to the new co-ordinates which, in contrast to the spherical ones,
have no sense of values measured by rulers and clock. After that, of course, 
we obtain the same
physical results as in the spherically co-ordinate.
( Like classical electrodynamics in
arbitrary co-ordinates). Therefore, since at $\mathcal{K=Q}$ the solution of
our equation have no the singularity in the center and events horizon, it 
does
 not contain theirs and in others coordinate systems.

We can argue that the constant 
${\mathcal{Q}}=r_{g}$. Indeed, consider the
00-component of eq. (\ref{Vr_EqPsiMatter}). Let us set
$T_{\alpha\beta}=\rho c^{2}u_{\alpha}u_{\beta },$ where $\rho$ is
the matter density and $u_{\alpha}$ is the 4-velocity of matter
points. At the small macroscopic velocities of the matter we can
set $u_{0}=1$ and $u_{i}=0$. Therefore, the equation is of the form%

\begin{equation}
\square\psi_{\alpha\beta}=\chi(\rho c^{2}+t_{00})
\end{equation}
where $\chi=4 \pi G/c^{4}$ and $t_{00}$ is the 00-component of the tensor%

\begin{equation}
t_{\alpha\beta}=\chi^{-1}B_{\alpha\sigma}^{\gamma}B_{\beta\gamma}^{\sigma}.
\label{Vr_TensorEnergyGravField}%
\end{equation}

Let us find the energy of a gravitational field of the point mass $M$ as the
following integral in the pseudo - Euclidean space - time
\begin{equation}
{\mathcal{E}}=\int t_{00}dV, \label{Vr_energydef}%
\end{equation}
resulting from the above solution, where $dV$ is the volume
element. In the Newtonian theory this integral is divergent. In
our case we have:
\begin{equation}
t_{00}=2\chi^{-1}B_{00}^{1}\;B_{01}^{0}=\frac{c^{4}}{8\pi G}\frac
{{\mathcal{Q}}^{2}}{{\mathcal{K}}} \label{Vr_t00}%
\end{equation}
and, therefore, in the spherical coordinates, we obtain \ %

\begin{equation}
{\mathcal{E}}=\int
t_{00}dV=\frac{1}{8}\frac{{\mathcal{Q}}^{2}c^{4}}{\pi G}J,
\label{Vr_energycalc}%
\end{equation}
where%

\begin{equation}
J=\int\frac{dV}{f^{4}}=\frac{4\pi}{3{\mathcal{K}}}B(1,1/3)
\end{equation}
and
\begin{equation}
B(z,w)=\int_{0}^{\infty}\frac{t^{z-1}}{(1+t)^{z+w}}dt \label{Vr_B-function}%
\end{equation}
is B-function. Using the equality
\begin{equation}
B(z,w)=\frac{\Gamma(z)\Gamma(w)}{\Gamma(z+w)}, \label{Vr_BataGamma}%
\end{equation}
where $\Gamma$ is $\Gamma$-function we obtain
$J=4\pi/{\mathcal{K}}$, and, therefore,
\begin{equation}
{\mathcal{E}}=\frac{{\mathcal{Q}}}{{\mathcal{K}}}
\frac{{\mathcal{Q}}c^{4}}{2G}%
=\frac{{\mathcal{Q}}}{{\mathcal{K}}}M\,c^{2} \label{Vr_energyfinally}%
\end{equation}

We arrive at the conclusion that at ${\mathcal{K}}\neq0$ the
energy of the point mass is finite and \ at
$\mathcal{K}=\mathcal{Q}$ the rest energy of the point
particle in full is caused by its gravitational field:%

\[
{\mathcal{E}}=M\,c^{2}.
\]

The spacial components of the vector $P_{\alpha}=t_{0\alpha}$ are equal to 
zero.

Due to these facts we assume in the present paper that ${\mathcal{K}}%
={\mathcal{Q}}=r_{g}$ and consider solution
(\ref{Vr_ABCequal}) in the spherical coordinates system at the
used gauge condition as a basis for the subsequent analysis.

\section{Orbits of Non-Zero Mass Particles.}

The equations of motion of a test particle of a non-zero mass in
the spherically symmetric field resulting from eqs.
(\ref{Vr_MyEqVacuum}) are given by \cite{Vr_Verozub91}
\begin{equation}
{\overset{\cdot}{r}}^{2}=(c^{2}C/A)[1-(C/\overline{E})(1+r_{g}^{2}
\overline{J}^{2}/B)],
\label{Vr_EqsMotionTestPart1}%
\end{equation}%
\begin{equation}
\overset{\cdot}{\varphi}=c\;C\overline{J}r_{g}/(B\overline{E}) 
\label{Vr_EqsMotionTestPart2}%
\end{equation}
where $(r,\varphi,\theta)$ are the spherical coordinates ($\theta$
is supposed to be equal to $\pi/2$), $\overset{\cdot}{r}=dr/dt$,
$\overset{\cdot}{\varphi}=d\varphi/dt$ ,
$\overline{E}=E/(mc^{2})$, $\overline{J}=J/(amc)$, $E$ is the
particle energy, $J$ is the angular momentum. Let
$\overline{u}=1/\overline{f}$, where
$\overline{f}=(1+\overline{r}^{3})^{1/3}$ and
$\overline{r}=r/r_{g}$. Then the differential equation of the
orbits, following from eqs. (\ref{Vr_EqsMotionTestPart1}) and
(\ref{Vr_EqsMotionTestPart2}) can be written as
\begin{equation}
(d\overline{u}/d\varphi)^2 = {\mathcal{G}}(\overline{u}) 
\label{Vr_DifEqOrbPart}%
\end{equation}
where
\[
{\mathcal{G}}(\overline{u})=\overline{u}^{3}-\overline{u}^{2}+\overline{u}/
\overline{J}^{2}+(\overline{E}^{2}-1)/\overline{J}^{2}
\]

eq. (\ref{Vr_DifEqOrbPart}) differs from the orbit equations of
general relativity \cite{Vr_Chandrasekhar} by the function
$\overline{f}$ instead of the function $\overline{r}$. Therefore,
the distinction in the orbits becomes apparent only at the
distances $\overline{r}$ of the order of 1 or less than that.

\begin{figure}[tbh]
\centering\noindent
\includegraphics[width=7cm,height=6cm]{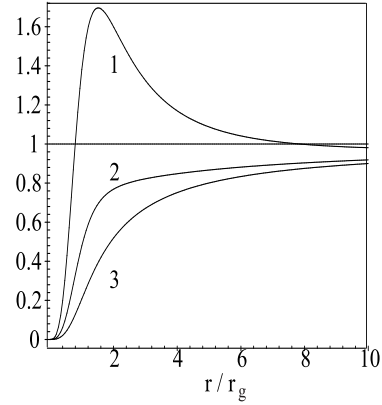}
\caption{The effective potential
${\mathcal N}={\mathcal N}(\overline{r})$ of the particles with
$\overline{J}=3,3^{1/3},0.$ (Curves 1, 2 and 3, correspondingly).}%
\label{Vr_ef_pot_part}%
\end{figure}

Setting $\overset{\cdot}{r}=0$ in eq.
(\ref{Vr_EqsMotionTestPart1}) we obtain $\overline
{E}^{2}={\mathcal N}(\overline{r})$, where
\begin{equation}
{\mathcal N}(\overline{r})=(1-1/\overline{f})
(1+\overline{J}^{2}/\overline{f}^{2})
\label{Vr_EffPotentialPart}%
\end{equation}
is the effective potential \cite{Vr_Chandrasekhar}.
Fig.(\ref{Vr_ef_pot_part}) shows the function
${\mathcal N}={\mathcal N}(\overline{r})$.

The function ${\mathcal{N}}(\overline{r})$ differs from the one in
general relativity in two respects:\newline 1. It is defined at
every point of the interval $(0,\infty)$\newline 2. It tends to
zero when $\overline {r}\rightarrow0 $.

\begin{figure}[htb]
\centering\noindent
\includegraphics[height=5cm,width=5cm]{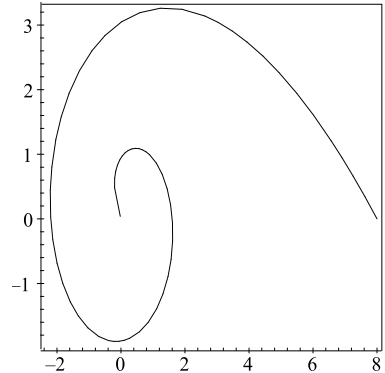}
\caption{The orbit of the particle with $\overline{J}=1.99$ and
$\overline{E}=1$.}
\label{Vr_orbit1_part}
\end{figure}

Possible orbit types can be shown by the horizontals
$\overline{E}=Const$. Two types of the orbits have peculiarity in
comparison with the Einstein
equations. The horizontals placed above the maximum of the curve 
${\mathcal{N}}%
(\overline{r})$ show the particles orbits which begin in the field
center and end in the infinity. In other words, for each value of
$\overline{J}$ there exists such a value of $\overline{E}$ for
which the gravitational field cannot keep particles escaping from
the center. The events horizon is absent. Fig.\ref{Vr_orbit1_part}
shows an example of the orbits at $\overline{J}=1.99$ and
$\overline{E}=1$.

The horizontals placed between Y-axis and the curve
${\mathcal{N}}(\overline{r})$ can show particles orbit kept by the
gravitational field near the field center.

It follows from eqs. (\ref{Vr_EqsMotionTestPart1}) and
(\ref{Vr_EqsMotionTestPart2}) that the velocity of a test particle
freely falling to the point mass $M$ tends to zero when
$r\rightarrow0$. The time of the motion of the particle from some
distance $r=r_{0}$ to $r=0$ is infinitely large. We can say that
the spherically symmetric solution has no physical singularity.

The points of the minimum of the function
${\mathcal{N}}(\overline{r})$ show stable closed orbits, the points
of the maximum show instable ones. The minimum of the function
${\mathcal{N}}(\overline{r})$ exists only at $\overline
{J}>\sqrt[3]{3}$ which corresponds to the value of the function
$f(\overline {r})>3$. Therefore, stable circular orbits exist only
at $\overline
{r}>\overline{r}_{cr}$, where $\overline{r}_{cr}=\sqrt[3]{26}\;$ $r_{g}%
\approx2.96\;$ $r_{g}$. The orbital speed of the particle with
$r=r_{cr}$ is equal to $0.4c$. At $r<r_{cr}$ unstable circular
orbits can exist. At $J\rightarrow\infty$ the location of the
maximums tends to $\overline{f}=3/2$. Therefore, the minimum
radius of the instable circular orbit is $\overline
{r}_{min}=1.33$ $r_{g}$. (In general relativity it is equal to
$1.5$ $r_{g}$). The speed of the motion of a particle on this
orbit is equal to $0.51c$. The binding energy
$\overline{E}=0.0572$, just as it occurs in general relativity.

The rotation frequency $\omega_{r}=\overset{\cdot}{\varphi}$ of the circular 
motion will
be
\begin{equation}
\omega_{r}=[c(\overline{f}-1)/(\overline{f}^{3}r_{g})](\overline{J}%
/\overline{E}) \label{Vr_omegaParticle}%
\end{equation}

In a circular motion $\overline{r}$ is the constant and,
therefore, the function ${\mathcal{N}}(\overline{r})$ has the
minimum. Consequently, from the equation $d{\mathcal{N}}/dr=0$ we
find
\begin{equation}
\overline{J}^{2}=\overline{f}^{2}/(2\overline{f}-3) \label{Vr_J}%
\end{equation}
Using (\ref{Vr_EqsMotionTestPart1}) we have at
$\overset{\cdot}{r}=0$
\begin{equation}
\overline{E}^{2}=2(\overline{f}-1)/[\overline{f}(2\overline{f}-1)] 
\label{Vr_E}%
\end{equation}

Equations (\ref{Vr_EffPotentialPart})$\div$ (\ref{Vr_J}) yield
\[
\omega_{r}=r_{g}^{1/2}c/(\sqrt{2}f^{3/2})
\].
Hence, the circular orbits have the rotation period
\begin{equation}
T=2\sqrt{2}\pi c^{-1}r_{g}(1+\overline{r}^{3})^{1/2} 
\label{Vr_Tpart}%
\end{equation}
( 3 rd Kepler law). In comparison with general relativity the
change in $T$ is $2\%$ at $\overline{r}=3$ and $20\%$ at
$\overline{r}=1.33$.

Consider the apsidal motion. For ellipsoidal orbits the function
${\mathcal{G}}(\overline{u})$ has 3 real roots
$\overline{u}_{1}<\overline {u}_{2}<\overline{u}_{3}$
\cite{Vr_Chandrasekhar}. The apsidal motion per one period is
\begin{equation}
\delta\varphi=2|\Delta|-2\pi, \label{Vr_apsMotion}%
\end{equation}
where
\begin{equation}
\Delta=\int_{\overline{u}_{1}}^{\overline{u}_{2}}[{\mathcal{G}}
(\overline{u})^{1/2}%
d\overline{u}. \label{Vr_Delta}%
\end{equation}

Consequently, \cite{Vr_Byrd}
\begin{equation}
\Delta=2(\overline{u}_{3}-\overline{u}_{1})^{1/2}{\mathcal{F}}(\pi/2,q),
\label{Vr_Delta1}%
\end{equation}
where
\begin{equation}
q=[(\overline{u}_{2}-\overline{u}_{3})/(\overline{u}_{3}-\overline{u}%
_{1})]^{1/2} \label{Vr_q}%
\end{equation}

and
\begin{equation}
{\mathcal{F}}(\pi/2,q)=\int_{0}^{\pi/2}(1-q^{2}\sin^{2}(\beta))^{1/2}d\beta.
\label{Vr_CalF}%
\end{equation}

Let us introduce (by analogy with general relativity) the following
notations:
\begin{equation}
\overline{u}_{1}=(1-e)/\overline{p},\;\overline{u}_{2}=(1+e)/\overline
{p},\;\overline{u}_{3}=1-2/\overline{p}, \label{Vr_u1u2u3}%
\end{equation}
where the parameter $\overline{p}$ at $\overline{r}\rightarrow\infty$ becomes
the focal parameter $p$ divided by $r_{g}$.

At $\overline{r}\gg1$ the value of $q\approx(2\overline{e}/\overline{p}%
)^{1/2}\ll1$ and, therefore,
\begin{equation}
{\mathcal{F}}(\pi/2,q)=(\pi/2)(1+q^{2}/4+9q^{2}/64+... ). \label{Vr_Fapprox}%
\end{equation}

Using eqs. (\ref{Vr_Delta}) and (\ref{Vr_CalF}) we find with
accuracy up to $1/\overline{p}^{2}$
\begin{equation}
\Delta\varphi=3\pi/\overline{p}+(\pi/(8\overline{p}^{2})(-e^{2}+16e+54+...).
\end{equation}

For the orbits of Mercury or a binary pulsar (such as PSR 1913 + 16 ) the
value of $\overline{u}$ differs very little from the value of $r_{g}/r$.
Consequently, the values of $\overline{p}=2/(\overline{u_{1}}+
\overline{u_{2}%
})$ and $e=(\overline{u_{2}}-\overline{u_{1}}/(\overline{u_{2}}+\overline
{u_{1}})$ differ very little from the values of $p/$ $r_{g}$ and the $e$.
Hence, their apsidal motion differs very little from the general relativity
prediction. Even, for example, at $\overline{p}=10$ and $e=0.5$, the
difference in $\Delta\varphi$ is about $6\cdot10^{-4}rad/year$.

\section{ Photon Orbits.}

The equations of motion of a photon in the spherical symmetric
field are given by \cite{Vr_Verozub91}
\begin{equation}
\overset{\cdot}{r}^{2}=(c^{2}\;C/A)(1-C\;b^{2}/B),\;\;
\overset{\cdot}{\varphi}=cCb/B,
\label{Vr_eqsMotionPoton}%
\end{equation}
where $b$ is an impact parameter.

The differential equation of the orbits can be written as
\begin{equation}
(d\overline{u}/d\varphi)^2 = {\mathcal{G}}_{1}(\overline{u}), 
\label{Vr_difEqsMotionPhoton}%
\end{equation}
where
\[
{\mathcal{G}}_{1}(\overline{u})=\overline{u}^{3}-\overline{u}^{2}+
1 / \overline{b}^{2}%
\]
and $\overline{b}=b/$ $r_{g}$.

\begin{figure}[htb]
\centering\noindent
\includegraphics[height=6cm,width=6cm]{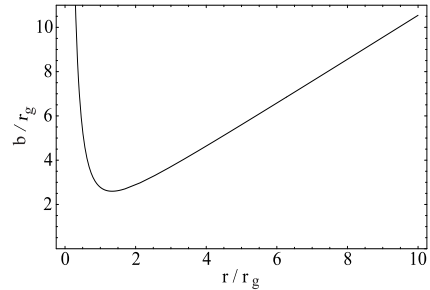}
\caption{The impact parameter $\overline{b}$ as the function of 
$\overline{r}$}
\label{Vr_ef_pot_phot}
\end{figure}

Setting $\overset{\cdot}{r}=0$ in
eq. (\ref{Vr_difEqsMotionPhoton}) we obtain 
$b=\overline{f}/(1-1/\overline{f})^{1/2}$ where
$\overline {f}=f/r_{g}$. Fig \ref{Vr_ef_pot_phot} shows 
$b$ as a function of $\overline{r}$, i.e. the location
of the orbits turning points.

The function $b(\overline{r})$
 is defined at all $r > 0$. The minimal value 
of $b$,
i.e. $b_{min}$, is equal to $2.6$. It is reached at
$\overline{r}_{min}=1.33$. The motion of photons can be shown by
the horizontal $\overline{b}=Const$. There are two types of 
orbits which have the peculiarities due to the lack of the events
horizon. The curves placed below the minimum of the curve
$\overline{b}(\overline{r})$ show that the attracting mass cannot
keep a photon escaping from the center at the parameter
$\overline{b}<\overline{b}_{min}$.The orbits of this type also
show the gravitational capture of the photon. The photon finishes
at the field center, unlike general relativity, where it ends on
the Schwarzshild sphere.

The angle of the light deflection at the distances close to $\overline
{r}=1.33$ is given by
\begin{equation}
\theta=\ln(4.021/\delta^{2}), \label{Vr_theta}%
\end{equation}
where $\delta=2(\overline{f}_{min}-3/2)$ and $\overline{f}_{min}%
=(1+\overline{r}_{min}^{3})^{1/3}$. It differs very little from
the one in general relativity in a weak field. (See also
\cite{Vr_Verozub2} ).


\section{Spherical-symmetric solution in ge\-ne\-ral case}

Properties of the gravitational field, by definition, display
themselves exclusively owing to its influence on the motion of test
particles. Let us show that the motion of test particles described
by  Lagrangian (\ref{Vr_LagrangianThirr}) is insensitive to the
fact whether $A, B, C$ are depending of time , or not.

Suppose, in the metric form (\ref{Vr_ds}) the coefficients $A$, $B$, $C$
are the functions of $r$ and $x^{0}$. The Christoffel symbols for
this form are given by 
\begin{align*}
\Gamma_{tt}^{t}   =\frac{1}{2}\frac{\overset{\cdot}{C}}{C};\;\;\;
\Gamma_{tt}^{r}%
=\frac{1}{2}\frac{C^{\prime}}{A};\;\;\;
\Gamma_{rr}^{r}   =\frac{1}{2}\frac{A^{\prime}}{A};\\
\Gamma_{rr}^{t}=\frac{1}{2}\frac{\overset{\cdot}{A}}{C};\;\;\;
\Gamma_{r\theta}^{\theta}   =\frac{1}{2}\frac{B^{\prime}}{B};\;\;\;
\Gamma
_{r\varphi}^{\varphi}=\frac{1}{2}\frac{B^{\prime}}{B};\\
\Gamma_{rt}^{r}   =\frac{1}{2}\frac{\overset{\cdot}{A}}{A};\;\;\;
\Gamma_{rt}^{t}%
=\frac{1}{2}\frac{C^{\prime}}{C};\;\;\;
\Gamma_{\theta\theta}^{r}=-\frac{1}{2}\frac{B^{\prime}}{A};\\
\Gamma_{\theta\theta}^{t}=\frac{1}{2}\frac{\overset{\cdot}{B}}{C};\;\;\;
\Gamma_{\theta\varphi}^{\varphi}  =\frac{\cos \theta  }
{\sin \theta};\;\;\;
\Gamma_{\varphi t}^{\varphi} =\frac{1}{2}\frac{\overset{\cdot}{B}}{B}\\
\Gamma_{\theta
t}^{\theta}=\frac{1}{2}\frac{\overset{\cdot}{B}}{B};\;\;\;
\Gamma_{\varphi\varphi}^{r}   =-\frac{1}{2}\frac{B^{\prime}\sin
^{2}\theta }{A};\;\;\;\\
\Gamma_{\varphi\varphi}^{\theta}   =-\sin \theta\;  \cos\theta ;
\Gamma_{\varphi\varphi}^{t}   =\frac{1}{2}\frac{\overset{\cdot}{B}\sin^{2}
\theta }{C};
\end{align*}
In these formulas we denote by prime the partial derivative with respect to
$r$ and by point - the partial derivative with respect to $x^{0}.$ Then the
equations of the motion can be written in the form

\begin{eqnarray}
\frac{d^{2}x^{0}}{ds^{2}}+\frac{\overset{\cdot}{C}}{2C}\left(\frac{dx^{0}}
{ds}\right)^{2}+\frac{C^{\prime}}{C}\frac{dx^{0}}{ds}\frac{dr}{ds}
\nonumber \\
+\frac{\overset{\cdot}{A}}{2C}\left(\frac{dr}{ds}\right)^{2}  
+\frac{\overset{\cdot}{B}}{2C}\left(\frac{d\varphi}{ds}\right)^{2}   =0,
\label{Vr_x0EquGeodesic}
\end{eqnarray}

\begin{eqnarray}
\frac{d^{2}r}{ds^{2}}+\frac{\overset{\cdot}{A}}{2A}\frac{dx^{0}}{ds}
\frac{dr}{ds}
+\frac{A^{\prime}}{2A}\left(  \frac{dr}{ds}\right)  ^{2}+
\frac{\overset{\cdot}{A}}
{2A}\frac{dx^{0}}{ds}\frac{dr}{ds} \nonumber \\
 - \frac{B^{\prime}}{2A}\left(  \frac
{d\varphi}{ds}\right)  ^{2} +\frac{C^{\prime}}{2A}\left(
\frac{dx^{0}}{ds}\right)^{2}=0,\;\;\; \label{Vr_rEquGeodesic}
\end{eqnarray}%

\begin{eqnarray}
\frac{d^{2}\theta}{ds^{2}}+\frac{\overset{\cdot}{B}}{B}\frac{dx^{0}}{ds}
\frac{d\theta
}{ds}-\sin\left(  \theta\right)  \cos\left(  \theta\right)  \left(
\frac{d\varphi}{ds}\right)  ^{2} \nonumber \\
+\frac{B^{\prime}}{2B}\frac{dr}{ds}\frac{d\theta}{ds}=0,\;\;\;
\label{Vr_thetaEquGeodesic}%
\end{eqnarray}%

\begin{eqnarray}
\frac{d^{2}\varphi}{ds^{2}}+\frac{\overset{\cdot}{B}}{B}\frac{dx^{0}}{ds}
\frac{d\varphi
}{ds}+\frac{B^{\prime}}{B}\frac{dr}{ds}\frac{d\varphi}{ds}=0.
\label{Vr_phiEquGeodesic}%
\end{eqnarray}
In these equations the prime and point denote the partial derivatives with
respect to $r$ and $dx^{0}$, respectively.

Now let us find the motion integrals.

Equation (\ref{Vr_thetaEquGeodesic}) has the solution
$\theta=\pi/2 .$ For this reason it can be assumed that the orbits
are in the plane $\theta=\pi/2
.$ Equation (\ref{Vr_phiEquGeodesic}) can be written in the form%

\[
\frac{d^{2}\varphi}{ds^{2}}+\frac{1}{B}\frac{dB}{ds}\frac{d\varphi} {ds}=0,
\]
or%

\begin{equation}
\frac{d}{ds}\left[  \ln\left(  \frac{d\varphi}{ds}\right)  +\ln\left(
B\right)  \right]  =0, \label{Vr_phiEqu1}%
\end{equation}
where%

\[
\frac{dB}{ds}=\frac{\partial B}{\partial x^{0}}\frac{dx^{0}}{ds}%
+\frac{\partial B}{\partial r}\frac{dr}{ds}%
\]
is the total derivative of the function $B\left(  r,x^{0}\right)  $ with
respect to $s$ along the world line of the particle. It yields the first
integral of the motion:%

\begin{equation}
B\frac{d\varphi}{ds}=J, \label{Vr_1IntegralOfMotion}%
\end{equation}
where $J$ is a constant.

Using the relation%

\begin{equation}
\overset{\cdot}{A}\left(  \frac{dr}{ds}\right)  ^{2}+\overset{\cdot}{B}
\left(  \frac{d\varphi}%
{ds}\right)  ^{2}-\overset{\cdot}{C}\left(  \frac{dx^{0}}{ds}\right)  ^{2}=0,
\label{Vr_1Equation}%
\end{equation}
that follows from the identity%

\begin{eqnarray}
C\left(  \frac{dt}{ds}\right)  ^{2}-A\left(  \frac{dr}{ds}\right)
^{2}-B\left(  \frac{d\theta}{ds}\right)  ^{2}-\\ B\sin^{2}\left(
\theta\right)
\left(  \frac{d\varphi}{ds}\right)  ^{2}=1 \label{Vr_2Equation}%
\end{eqnarray}
at $\theta=\pi/2,$ equation (\ref{Vr_x0EquGeodesic}) can be transformed to 
the form%

\begin{equation}
\frac{d^{2}x^{0}}{dt^{2}}+\frac{1}{C}\frac{dC}{ds}\frac{dx^{0}}{ds}=0,
\label{Vr_3Euation}%
\end{equation}
where $dC/ds$ is the total derivative of the function $C\left(  r,x^{0}%
\right)  $ with respect to $s$. That yields the second integral of the 
motion%

\begin{equation}
C\frac{dx^{0}}{ds}=\lambda. \label{Vr_2IntegralOfMotion}%
\end{equation}
where $\lambda$ is a constant. eq. (\ref{Vr_rEquGeodesic}), after being
multiplied by
the factor $2 A dr / ds$ can be written in the form%

\begin{eqnarray}
 \frac{d}{ds}\left[  A\left(  \frac{dr}{ds}\right)  ^{2}\right]  +
 \overset{\cdot}
{A}\frac{dx^{0}}{ds}\left(  \frac{dr}{ds}\right)  ^{2} \nonumber  \\
+C^{\prime}\left(\frac{dx^{0}}{ds}\right)  ^{2}\frac{dr}{ds}-
 B^{\prime}\left(  \frac{d\varphi}{ds}\right)  ^{2}\frac{dr}{ds}%
=0;\label{Vr_4Equation}%
\end{eqnarray}
and, after that, by using relation (\ref{Vr_1Equation}), in the form%

\begin{equation}
\frac{d}{ds}\left[  A\left(  \frac{dr}{ds}\right)  ^{2}\right]  -\frac{dB}%
{ds}\left(  \frac{d\varphi}{ds}\right)  ^{2}+\frac{dC}{ds}\left(  \frac
{dx^{0}}{ds}\right)  ^{2}=0, \label{Vr_5Equation}%
\end{equation}
where $dB/ds$ is the total derivative of the function $B\left(  r,x^{0}%
\right)  $. Taking into account (\ref{Vr_1IntegralOfMotion}) and
(\ref{Vr_2IntegralOfMotion}), we obtain%

\begin{equation}
\frac{d}{ds}\left[  A\left(  \frac{dr}{ds}\right)  ^{2}+J^{2}\frac{d}%
{ds}\left(  \frac{1}{B}\right)  -\lambda^{2}\frac{d}{ds}\left(  \frac{1}%
{C}\right)  \right]  =0, \label{Vr_6Equation}%
\end{equation}
which yields the third integral of the motion:%

\begin{equation}
A\left(  \frac{dr}{ds}\right)  ^{2}-\frac{\lambda^{2}}{C}+\frac{J^{2}}{B}=E,
\label{Vr_3IntegralOfMotion}%
\end{equation}
where $E$ is a constant.

Now it can be demonstrated that the same integrals of the motion
can be obtained from Lagrangian (\ref{Vr_LagrangianThirr}) for
the static field. At
$\theta=\pi/2$ it can be written in the form%

\begin{equation}
L =-m_{p}c^{2}\left(  c^{2} C - A \overset{\cdot}{r}^{2}-B \overset{\cdot}{
\varphi}^{2}\right)
^{1/2}, \label{Vr_LagrangianThirr1}%
\end{equation}
where $A,B$ and $C$ are the functions of $r$ only. Since $L$ does not depend
on $x^{0}$ and $\varphi$, there are two integrals of the motion:%

\begin{equation}
\overset{\cdot}{r}\frac{\partial L}{\partial\overset{\cdot}{r}}+\overset{
\cdot}{\varphi}\frac{\partial
L}{\partial\overset{\cdot}{\varphi}}-L=Const; \label{Vr_EquationOfLagrang1}%
\end{equation}%

\begin{equation}
\frac{\partial L}{\partial\overset{\cdot}{\varphi}}=Const. 
\label{Vr_EquationOfLagrang2}%
\end{equation}
The first of these equations is of the form%

\[
\frac{\overset{\cdot}{r}^{2}A+\overset{\cdot}{\varphi}^{2}B}{L}+
\frac{L}{m_{p}^{2}c^{4}}=Const.
\]
But it follows from eq. (\ref{Vr_LagrangianThirr1}) that%

\begin{equation}
\frac{A\overset{\cdot}{r}^{2}+B\overset{\cdot}{\varphi}^{2}}{L}= 
\frac{C}{L}-\frac{L}{m_{p}^{2}%
c^{4}}. \label{Vr_1Lagrang}%
\end{equation}
Consequently, $C/L=Const.$ However,%

\begin{equation}
-\frac{L}{m_{p}c^{2}}L=\frac{ds}{dx^{0}} \label{Vr_2Lagrang}%
\end{equation}
Thus, we arrive at equation (\ref{Vr_2IntegralOfMotion}).

The second integral of the motion can be found from  equations
(\ref{Vr_EquationOfLagrang2}) and (\ref{Vr_2Lagrang}) and
coincides with (\ref{Vr_1IntegralOfMotion}).

To obtain the third integral of the motion we start from the identity%

\begin{equation}
g_{\alpha\beta}\frac{dx^{\alpha}}{ds}\frac{dx^{\beta}}{ds}=\zeta,
\label{Vr_Curvature}%
\end{equation}
where $\zeta$ or the particle with the mass $m_{p}\neq0$ and
$\zeta=0$ for the particles with the mass $m_{p}=0.$ Substituting
(\ref{Vr_Curvature}) $d\varphi/ds$ and $dx^{0}/ds$ from 
eqs. (\ref{Vr_1Lagrang}) and (\ref{Vr_2Lagrang}) into this
equation, we obtain the equality which coincides with 
eq. (\ref{Vr_3IntegralOfMotion}).

Since the properties of a gravitation field are defined by their
influence on the motion of the particles, a solution of any
correct gravitation equations for the spherical symmetric field,
based on Lagrangian (\ref{Vr_LagrangianThirr1}), must be static.

Using the above results, we can also find gravitational field
inside a spherically - symmetric matter layer. In order to reach a
coincidence of the motion equations of the test particles in the
nononrelativistic limit with the Newtonian ones, the constant
$r_{g}$ in eq. (\ref{Vr_Cequal}) in that case must be set equal to
zero. Therefore, the spherically - symmetric matter layer does not
create the gravitational field inside itself.

\section{ Equilibrium Configurations with Large Masses}

In this Section we consider one of the most interesting
consequences of gravitation equations (\ref{Vr_MyEqVacuum}) -
the possibility of the existence of compact configurations of
degenerated Fermi- gas with very large masses.

The radial component of the gravity force affecting a test
particle at rest in the spherically - symmetric field is given by
\cite{Vr_Verozub91}
\begin{equation}
F^{r}=-mB_{00}^{r}=-\frac{G\,m_{p}\,M}{r^{2}} \left[  1-\frac{r_{g}}
{(r^{3}+r_{g}^{3})^{1/3}} \right]  \label{Vr_F}%
\end{equation}

\begin{figure}[htb]
\centering\noindent
\includegraphics[height=6cm,width=7cm]{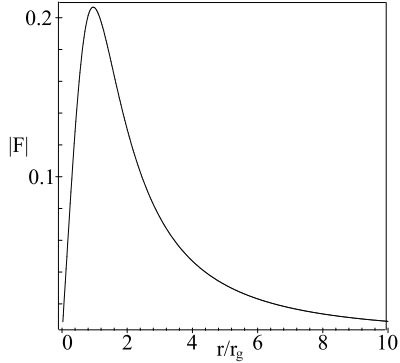}%
\caption{The modulus of the gravitational force (arbitrary units) as the 
function of the
distance from the center ($r/r_{g}$).} 
\label{ForceStat}
\end{figure}

It follows from this figure that  $|F|$ reaches its maximum at the distance
$r$ of the order of $r_{g}$ and tends to zero at $r\rightarrow0$.

It would therefore be interesting to know what masses of the
equilibrium configurations can exist if gravitational force is given by
eq. (\ref{Vr_F}).
To answer this
question we start from the equation%

\begin{equation}
\frac{dp}{dr}=-\frac{G\rho m}{r^{2}}\left[  1-\frac{r_{g}}{(r^{3}+r_{g}%
^{3})^{1/3}}\right]  .\label{Vr_dp/dr=}%
\end{equation}

In this equation $p$ is the pressure, $m=m(r)$ is the matter mass inside a
sphere of the radius $r$, $\rho=\rho(r)$ is the matter density at the 
distance
$r$ from the center, $r_{g}$ is the function of $m(r)$.

Suppose the equation of state is $p=K\rho^{\Gamma}$, where
$K$ and $\Gamma$ are constants. For numerical estimates we shall
use their values from  \cite{Vr_Shapiro}.

For rough estimates we set $\rho=Const$ and replace $dp/dr$ by
$-p/R$, where $p$ is the average matter pressure and $R$ is its
radius. Under the circumstances we obtain from eq. 
(\ref{Vr_dp/dr=})
\begin{equation}
\frac{p}{\rho c^{2}}=\frac{r_{g}}{2R}\left(  1-\frac{r_{g}}{f}\right)
.\label{Vr_consequenceEqs2}%
\end{equation}

If $R\gg$ $r_{g}$, then the term $r_{g}/f$ is negligible. Setting
$M\approx\rho R^{3}$, we find the mass of equilibrium states as a
function of $\rho$:
\begin{equation}
M=(K/G)^{3/2}\rho^{(\Gamma-4/3)(3/2)}. \label{Vr_M1=}%
\end{equation}\sloppy
It yields the maximum Chandrasekhar mass \cite{Vr_Chandrasekhar}
$M$ $=$ $(K/G)^{3/2}$ at $\rho\gg\rho_{0}$.

However, according to eq. (\ref{Vr_consequenceEqs2}), there are
also equilibrium configurations at $R<$ $r_{g}$. In particular, at
$R\ll$ $r_{g}$ we find from eq.  (\ref{Vr_consequenceEqs2}) that
the masses of the equilibrium configurations are given by
\begin{equation}
M=c^{9/2}10^{-1}K^{-3/4}G^{-3/2}\rho^{-(\Gamma-1/3)(3/4)}. \label{Vr_M2=}%
\end{equation}
These are the configurations with very large masses. For example, the
following equilibrium configurations can be found:

the nonrelativistic electrons: $\rho=10^{5}$ $g/cm^{3}$, 
$M=1.3\cdot10^{42}$
$g$, $R=2.3\cdot10^{12}cm$,

the relativistic electrons: $\rho=10^{7}$ $g/cm^{3}$, 
$M=2.3\cdot10^{40}$ $g$,
$R=1.3\cdot10^{11}cm$,

the nonrelativistic neutrons: $\rho=10^{14}$ $g/cm^{3}$ $M=3.9\cdot10^{35}$
$g$, $R=1.6\cdot10^{7}cm$.

\begin{figure}[htb]
\centering\noindent
\includegraphics[height=5cm,width=6cm]{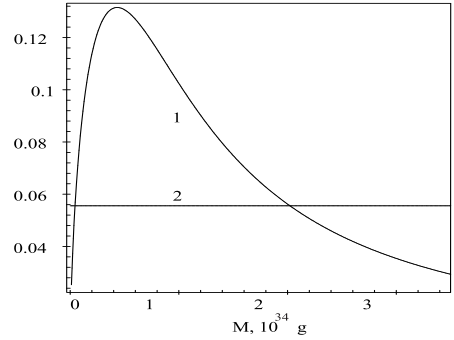}%
\caption{The plots of
right-hand and left-hand sides of eq. 
(\ref{Vr_consequenceEqs2}) against mass $M$}
\label{Vr_equilibrium1}
\end{figure}

The reason of the two types of configurations existence can be
seen from fig. \ref{Vr_equilibrium1}, where for $\rho=10^{15}\ $
$g/cm^{3}$, $K=5\cdot10^{9}$ and $\Gamma=5/3$ the plots of
right-hand and left-hand sides of eq. 
(\ref{Vr_consequenceEqs2}) against mass $M$ are shown.%

The following conclusions can be made after considering the plots of the 
above kind:

1. For each value of $\rho<\rho_{max}$ there are two equilibrium states 
(with
$R>$ $r_{g}$ and $R<$ $r_{g}$).

2. There are no equilibrium configurations whose density is larger than a
certain value $\rho_{max}\sim10^{16}$ $g/cm^{3}$. (At the densities 
exceeding
$\rho_{\max}$ the curves do not intersect).

More accurate conclusions about the internal structure of the
configuration can be obtained from the equation of the hydrostatic
equilibrium obtained for the gravity force (\ref{Vr_F}):
\begin{eqnarray}
\frac{dp}{dr}   =-\frac{\rho G\;m(r)}{r^{2}}\left[  1-\frac{r_{g}}%
{(r^{3}+r_{g}^{3})^{1/3})}\right]
,\label{Vr_hydrostatic_equations}\\ \frac{dm}{dr}  =4\pi\rho
r^{2},\nonumber
\end{eqnarray}
where $m=m(r)$ and $r_{g}=r_{g}[m(r)]$. The equation of state is
\begin{equation}
p=\left(  \frac{n_{b}}{dn/d\rho}-\rho\right)  \;c^{2},
\end{equation}
where the baryons density $n_{b}$ as the function of $\rho$ is
given by the approximation Harrison equation \cite{Vr_Harrison}
which takes place from $7$
to at least $10^{16}$ $g\,cm^{-3}$:%
\begin{equation}
n_{b}=A\rho(1+B\rho^{1/16})^{-4/9},\label{Vr_Harrison_state_equations}%
\end{equation}
where $A=6.0228\cdot10^{23}$ and $B=7.7483\cdot10^{-10}$ in CGS units.

\begin{figure}[htb]
\centering\noindent
\includegraphics[width=5cm,height=4cm]{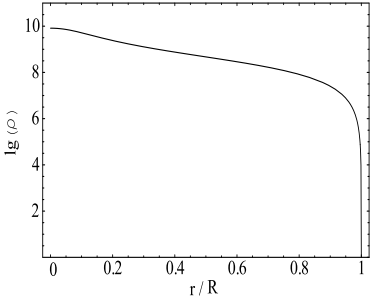}
\caption{The density distribution inside the object with the mass
$M=3.9\cdot 10^{35} g$ and the radius $R=1.6 \cdot 10^{7}cm$}
\label{Vr_density_of_r}
\end{figure}

In addition to the ordinary solution (i.e. configurations of the
white dwarfs and neutron stars) there exist solutions with large
masses. Fig. \ref{Vr_density_of_r} shows an example of that kind
of solutions for $\rho(r)$. It is a configuration with the mass
$2.6\cdot10^{6}M_{\odot}$ and the radius $0.378\cdot R_{\odot}$.

Are the configurations with large masses stable? The total energy of the
degenerate gas is $E=E_{int}+E_{gr}$, where $E_{int}$ is the intrinsic 
energy
and $E_{gr}$ is the gravitational energy. The gravitational energy of the
sphere with the radius $r$ is%

\begin{equation}
E_{gr}=-\int dm(r)\ \chi(r)\ m(r), \label{Vr_Egrav}%
\end{equation}
where
\[
\chi(r)=\int_{r}^{\infty}dr^{\prime}\ (r^{\prime})^{-2}(1-1/f),
\]
$r_{g}=2G\,m(r)/c^{2}$, $f=($ $r_{g}(r)^{3}+(r^{\prime})^{3})^{1/3}$,%

\begin{equation}
m(r)=4\pi\int_{0}^{r}dr^{\prime}\rho(r^{\prime})^{2} (r^{\prime})^{2}.
\label{Vr_m(r)}%
\end{equation}

The function $\chi(r)$ is given by%

\begin{equation}
\chi=\frac{1}{r_{g}\overline{r}}+\frac{1}{2r_{g}\overline{r}^{2}}{
\mathcal{F}}\left(
\frac{1}{2},\frac{2}{3};\frac{5}{3};-\frac{1}{\overline{r}^{3}}\right),
\end{equation}
where ${\mathcal{F}}$ is the degenerated hypergeometric function.
Approximately
\begin{equation}
\chi=(1/r)[(1-\exp(-0.7r/r_{g})].
\end{equation}
Therefore, at $p=Const$ up to a constant of the order one
\begin{equation}
E_{gr}=-\frac{G\,M^{2}}{R}\left [1-\exp(-0.7R/r_{g}) \right].
\label{Vr_EgrApprox}%
\end{equation}

\begin{figure}[htb]
\centering\noindent
\includegraphics[width=7.5cm,height=5cm]{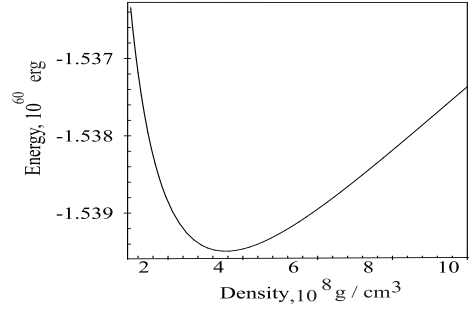}%
\caption{The plot of the energy $E$ vs. the density $\rho$ for the
configuration with the mass $M=2.5\cdot10^{6}$ $M_{\odot}$ }
\label{Vr_equilibrium2}
\end{figure}

\begin{figure}[htb]
\centering\noindent
\includegraphics[width=6cm,height=5cm]{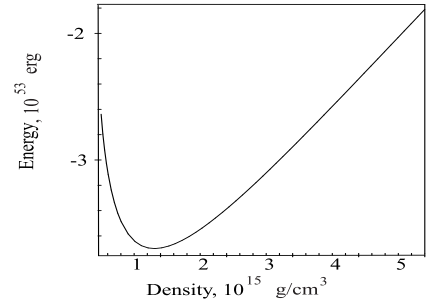}
\caption{The plot of the energy $E$ vs. the density $\rho$ for the
neutron star with mass $10^{33}$ $g$} \label{Vr_equilibrium3}
\end{figure}

The intrinsic energy $E_{int}=\int u\ dm$, where $u$ is the energy
per mass unit. For the used equation of state
$u=K(\Gamma-1)^{-1}\rho^{\Gamma-1}$. Thus, up to the constant of
the order of one
\begin{eqnarray}
E=KM\rho^{\Gamma-1}-  \nonumber \\ 
G\,M^{5/3}\rho^{1/3}
[1-\exp(-QM^{-2/3}\rho^{-1/3})],
\label{Vr_Efinal}%
\end{eqnarray}
where $Q=0.7c^{2}/2G$. As an example, fig.\ref{Vr_equilibrium2}
shows the plot of the function $E=E(\rho)$ for the nonrelativistic
degenerated Fermi gas of the mass $M=2.5\cdot10^{38}\ g$ in comparison
with the neutron star of the mass $M=10^{33}\ g$ in fig.
\ref{Vr_equilibrium3}.

The analysis of such plots shows that the function $E=E(\rho)$ has the 
minimum.
Thus, the above equilibrium states of the large masses are stable.

More rigorous investigation that confirms this result was carried out in 
\cite{VerKoch2001}.

\section{Conclusion}
It follows from the above results that the
equations under consideration do not contradict available
experimental data obtained in the Solar system. In paper
\cite{Vr_VerKoch2000} these equations were tested by the binary
pulsar PSR1913+16 and it was found out that the results are very close to
the ones in general relativity. It is a consequence of the fact
that the used distances from attracting masses are much larger
than the Schwarzshild radius. At the conditions the function $f(r)$
is very close to the radial distance $r$. 
However, the physical consequences
between these equations are completely different at the distances
$r\leq r_{g}$.
The events horizon is absent. There can
exist supermassive configurations of the degenerated Fermi-gas.
Candidates to the objects of such a kind are the galactic centers
(\cite{Vr_Eckart} $\div$ \cite{Vr_Bender}).


\begin{thebibliography}{99}                                                                                                %

\bibitem{Vr_Poincare}  H. Poincar\'{e}, Derni\`{e}res pens\'{e}es
Flammarion Paris (1913)
\bibitem{Vr_Berkley}  The Work of George  Berkeley.  Vol.2,
p.21-113;   Vol.4, p.11-30. London (1949)
\bibitem{Vr_Leibnitz}.  The Leibniz - Clarke - Correspondence. 
Manchester (1956)
\bibitem{Vr_Mach}  E. Mach, The Science of Mechanics. Open Court, La Salle,
1960)
\bibitem{Vr_Dehnen}  H. Dehnen, Wissensch. Zeitschr. der Fridrich - Schiller
Universitat.  Jena, Math. - Naturw. Reihe. H.1  Jahrg., 15, 15
(1966)
\bibitem{Vr_Landau} L. Landau and E.~Lifshitz, The Classical Theory of
Field. Addison - Wesley, Massacusetts (1971)
\bibitem{Vr_Rund}  H. Rund,  The differential geometry of Finsler
space. Springer (1959)
\bibitem{Vr_Post}  E. Post, {\it Rev. Mod. Phys.} {\bf 39},  475 (1967)
\bibitem{Vr_Ananden}  J. Ananden, {\it Phys. Rev. D.} {\bf 24}, 338 (1981)
\bibitem{Vr_Ashtekar}  A. Ashtekar and A.~Magnon, {\it Journ. of Math. Phys.} 
{\bf 16}, 341 (1975)
\bibitem{Vr_Shibata}  C. Shibata, H.~Shimada, M.~Azuma and H.~Yasuda, {\it
Tensor,} {\bf 31}, 219 (1977)
\bibitem{Vr_Syng}  J. L.~Syng, Classical dynamics. Springer-Verlag (1960)
\bibitem{Vr_Verozub} L.~V.~Verozub, {\it Ukr. Phys. Journ.} {\bf 26}, 
1598 (1981)
\bibitem{Vr_Verozub81} L. V.~Verozub, {\it Ukr. Phys.Journ.} {\bf 10}, 131, 
778, 1598
(1981)
\bibitem{Vr_Verozub95} L.~V.~Verozub, {\it Phys. Essays,} {\bf 8}, 518
(1995)

\bibitem{Vr_Rodichev} V. I.~Rodichev, In: Einstein Collection, 286 (1974)

\bibitem{Vr_Thirring} W. Thirring, {\it Ann. Phys.} {\bf 16}, 96 (1961)

\bibitem{Vr_Verozub91} L. V.~Verozub, {\it Phys. Lett. A,} {\bf 156},
404 (1991)

\bibitem{Vr_Levi-Chivita} T.~Levi-Chivita, {\it Ann. di Mat.}, Ser.2,
255 (1886)

\bibitem{Vr_Sinukov} N.~S.~Sinukov, Geodesic Mappings of Riemmanian Spaces.
Moscow, 255 (1979)

\bibitem{Vr_Petrov} A. Z.~Petrov, New Methods in General Relativity. 
Moscow (1966)

\bibitem{Vr_Berwald} L. Berwald, {\it Ann. of Math.}, {\bf 37}, 879
(1936)

\bibitem{Vr_Weyle} H. Weyl, {\it Bull. Amer. Math. Soc.}, {\bf 35},
716 (1929)

\bibitem{Vr_Whitehead} J. H.~Whitehead, {\it Ann. Math.}, {\bf 32},
327 (1931)

\bibitem{Vr_Weblen} O. Veblen, {\it Proc. Nat. Acad. Sci.}, {\bf 14},
154 (1928)

\bibitem{Vr_Schwarzschild} K. Schwarzschild, {\it d. Berl. Akad.},
189 (1916). [ Also in : "Albert Einstein and theory of gravitation".
Moscow, 199 (1979)]

\bibitem{Vr_Chandrasekhar} S. Chandrasekhar, The Mathematical Theory of
Black Holes. Oxford Univ. Press, New York (1983)

\bibitem{Abrams1} L. Abrams, {\it Phys.Rev.}, {\bf 20}, 2474 (1979)

\bibitem{Vr_VerKoch2001} L. V.~Verozub and A.~Y.~Kochetov, {\it Astr. 
Nachr.
}, {\bf 3}, 143 (2001)

\bibitem{Vr_VerKoch2000} L. V.~Verozub and A.~Y.~Kochetov, {\it Grav and
Cosmol.}, {\bf 6}, 246 (2000)

\bibitem{Vr_Byrd} P. F.~Byrd and M.~D.~Fridman, Handbook of elliptic
integrals for engineers and physicists. Berlin - Gottingen -
Heidelberg (1954)

\bibitem{Vr_Eckart} A. Eckart and  R.~Genzel, {\it MNRAS}, {\bf 284},
576 (1997)

\bibitem{Vr_Genzel}R. Genzel, D.~Hollenbach and C.~H.~Townes
{\it Rep. Progr. Phys.}, {\bf 57}, 417 (1994)

\bibitem{Vr_Chez} A. M.~Ghez, B.~L.~Klein, M.~Morris and E.~E.~Becklin,
{\it E-prepr. astro-ph/9807210}

\bibitem{Vr_Marel} R. P.~van der Marel, T.~de Zeeuw and H-W.~Rix,
{\it Ap. J.},{\bf 493}, 613 (1998)

\bibitem{Vr_Greenhill} L. Greenhill, M.~Nakai, P.~Diamond and M.~Inoue,
{\it Nature}, {\bf 373}, 127 (1995)

\bibitem{Vr_Ferrarese} L.~Ferrarese, H.~ C.~Ford and W.~Jaffe, 
{\it Ap. J.}, {\bf 470},
444 (1996)

\bibitem{Vr_Bender} R. Bender, J.~Kormendy and W.~Dehnen, {\it Ap. J.},
{\bf 464}, 123 (1996)

\bibitem{Vr_Verozub2} L.~ V.~Verozub, {\it Astr. Nachr.}, {\bf 317},
107 (1996)

\bibitem{Vr_Shapiro} S. Shapiro and S. Teukolsky, Black Holes,
White Dvarfs, and Neutron Stars. Jone Wiley \& Sons. (1983)

\bibitem{Vr_Harrison} B. K. Harrison, K.~ S.~Thorn, M.~Wakano,
J.~A.~Wheeler, Gravitational Theory and Gravitational Collapse.
Univ. of Chicago Press, Chicago, Illinois (1966)
\end{thebibliography}
\end{document}